\documentclass[aps,pra,amsmath,amssymb,preprint]{revtex4}
\usepackage{graphicx}
\usepackage{bm}
\def\d{{\rm d}}
\def\real{{\cal\rm R}{\rm e} \ }
\def\imag{{\cal \rm I}{\rm m} \ }
\def\BRA{\left\langle}
\def\KET{\right\rangle}
\def\LK{\left(}
\def\RK{\right)}
\def\LBK{\left\lbrack}
\def\RBK{\right\rbrack}
\def\LB{\left\lbrace}
\def\RB{\right\rbrace}
\def\Eq#1{(\ref{#1})}
\def\lv#1{\left( #1 \right\vert}
\def\rv#1{\left\vert  #1 \right)}

\def\DL{{L}}
\def\DQ{{\cal Q}}
\def\DP{{\cal P}}
\def\non{\nonumber}
\def\be{\begin{equation}}
\def\ee{\end{equation}}
\def\bea{\begin{eqnarray}}
\def\eea{\end{eqnarray}}

\def\Tr{\mathop{\rm Tr}}
\def\noi{\noindent}

\def\journal#1#2#3#4{{#1} {\bf #2}, #3 (#4)}
\def\PE{{\cal P}^{[1]}}
\def\QE{{\cal Q}^{[/1]}}
\begin{document}
\title{On the  modeling of irreversibility by relaxator Liouville dynamics}
\author{J\'anos Hajdu\footnote{passed away on September 20, 2025} and Martin Jan{\ss}en}
\email{mj@thp.uni-koeln.de}
\affiliation{Institut f\"ur Theoretische Physik, Universit\"at zu K\"oln, Z\"ulpicher Str. 77, D-50937 K\"oln}
\date{January 2, 2026 - revised version}
\begin{abstract}
A general approach to modeling irreversibility starting from microscopic reversibility is presented. The time $t_s$ up to which  relevant degrees of freedom of a system are tracked is extremely much shorter than the spectral resolution time $t_e$ that would be necessary to resolve the spectrum of all degrees of freedom involved. A relaxator that breaks reversibility condenses in the Liouville operator of the relevant degrees of freedom. The irrelevant degrees of freedom act as an environment to the system. The  irreversible relaxator Liouville equation contains memory effects and initial correlations of all degrees of freedom.   Stationary states  turn out to be generically unique and independent of the initial conditions and exceptions are due to degeneracies.   Equilibrium states  lie in the relaxator's kernel yielding a stationary Pauli master equation. Kinetic equations for one-particle densities are constructed as special cases of relaxator Liouville dynamics.  Kubo's linear response theory is generalized to relaxator Liouville dynamics  and related to irreversibility within the system. In a weak coupling approximation between system and environment the relaxator can be reduced to environmental correlations  and bilinear system  operators. Markov approximation  turns the relaxator Liouville dynamics into a semi-group dynamics.
\end{abstract}
\maketitle

\section{Introduction}\label{intro}
The microscopic von Neumann equation, $\dot{\rho}= -\imath L\rho$,  for quantum states $\rho$ is reversible for $L\cdot =\LBK H, \cdot\RBK$ with Hermitian Hamiltonian $H$. To each solution, $\rho(t)=e^{-\imath L t}\rho_0$, the inverse  operation, $e^{-\imath L (-t)}\rho(t) =\rho_0$, describes the reversed motion. The von Neumann entropy, $S=-\Tr \rho \ln \rho$ (we use units in which $\hbar=k_B=1$), of an arbitrary state $\rho$ stays constant according to the microscopic von Neumann equation (see e.g. \cite{Zwiebach2022}). Similar conclusions hold in classical Hamiltonian mechanics (Liouville theorem). 

Basic laws of thermodynamics  state that macroscopic systems consisting of a huge number $N$ of constituents reach a steady state after some characteristic relaxation time $\tau_r$. Lots of initial states reach the same stationary state. In closed systems with fixed average energy the stationary state is an equilibrium state, very well  described as a canonical Gibbs state, $e^{-H/T}/Z$ with temperature $T$ and partition sum $Z$.  Among all possible unit trace states with fixed average value of energy $E=\BRA H\KET$ the Gibbs state has the maximum von Neumann entropy.  In closed systems with fixed average energy $E$ the von Neumann entropy will change from its initial value to the maximum value of the Gibbs state. The reversed motion never happens. The condition of a macroscopic description by the Gibbs state as an equilibrium state relies on the condition of large $N$ manifesting in very small relative fluctuations of thermodynamic state variables like energy $E$ of the order of $1/\sqrt{N}$ (see e.g. \cite{LL5}).

The irreversibility of macroscopic processes expressed by the basic thermodynamic laws seems to be in contradiction with the reversibility of the microscopic von Neumann equation.  The debate of this issue goes back to the early days of statistical mechanics and is still with us (see e.g. \cite{Boltzmann1872}-\cite{Handbook2011}).
To stick to the von Neumann equation and modify e.g. the definition of entropy and/or  refer to the role of special initial conditions that are rare in some probabilistic  sense cannot suffice to recover macroscopic irreversibility, because reversible motions cannot relax to a steady state in a manner independent of the initial state. The dynamics should tell us about the relaxation time $\tau_r$ to reach stationarity. To extract relaxation as an emerging qualitative feature from the von Neumann equation it needs a treatment under the condition of macroscopic $N$. It is long known that microscopic reversibility has one  consequence which must be sacrificed for macroscopic irreversibility: recurrence of initial states after some recurrence time $T_R$. Thus, there is one obvious condition for macroscopic irreversibility, already formulated by Smoluchowski in 1916 \cite{Smoluchowski1916}: the available  time $t_s$ {{}  up to which  relevant degrees of freedom of a system are tracked} has to be negligible compared to the recurrence time $T_R$. {{}  To track relevant degrees of freedom in sufficient detail $t_s$ should be of the 
order of the typical inverse level spacing  of an isolated system of only relevant 
variables. It can thus be calculated as the density of frequency levels of an isolated 
system of only relevant degrees of freedom. $T_R$  can be calculated as the typical time after
which the return probability of an initial state reaches again a value significantly higher 
than its average value. For a time signal as superposition of oscillations with different periods, the time $T_R$ is the  common least multiplier of the periods.  A practical way of calculation is given in \cite{Ven2015}}. Typically, recurrence times grow double exponential with $N$ and quickly reach super-astronomic time scales for many body systems (for a detailed discussion see \cite{Ven2015}). 

As we will argue now, the essential condition for irreversibility in macroscopic systems is the smallness of $t_s$ with respect to the {{} spectral resolution time} $t_e$ that is needed to {{} resolve}  the full spectral properties of all degrees of freedom involved.  {{} $t_e$  can be calculated as the density of frequency levels of the total system or of an isolated system of only irrelevant degrees of freedom since the total system is assumed to
have a huge number of degrees of freedom almost all of which are irrelevant. By this it is the inverse of the typical spacing of eigenfrequencies of the total system in a certain dynamic regime of frequencies.}  In macroscopic systems $t_e$ scales exponentially with $N$ (see below) while $t_s$ does not depend on $N$. The condition  $t_s\ll t_e$ guarantees that the Smoluchowski criterion is fulfilled with ease. The condition $t_s\ll t_e$ means that we can treat the spectrum of $L$ as continuous with {{} a  continuous density of frequency levels} $\gamma_L(\Omega)$, because we cannot resolve {{} discrete levels} when following the system over reasonable  time scales of order $t_s$. One immediate consequence is that recurrence times are treated as infinite, because in a continuum of periods there is no finite common least multiplier. A second consequence corresponds to the relaxation of time dependent correlation functions $c_A(t)=\BRA A(t)A\KET $ of time dependent observables $A(t)$ with respect to a stationary state $\rho$,  $c_A(t)=\sum_n P^{[A]}_n e^{\imath \Omega_n t}$, with $P^{[A]}_n > 0$.  Such superposition with lots of different frequencies leads to almost cancellations and thus irreversible relaxation to a stationary value in a macroscopic system under a limited spectral  resolution. With full spectral resolution of all terms in the superposition, one could find almost recurrence of some  initial value after $T_R$, but on time scales $t_s\ll t_e\ll T_R$ the relaxation to stationarity is the appropriate picture \cite{Ven2015}. Indeed, to take imperfect resolution into account one approximates   the sum in $c_A(t)$  with an integral over the density $\gamma_L(\Omega)$, with $\gamma_L(\Omega)\cdot P^{[A]}(\Omega)$  of  a finite range of order $1/\tau_A$. Then one finds the correlations to relax to stationarity on a scale of order $\tau_A$. Thus, the condition of $t_s\ll t_e$ in macroscopic systems is the reason for treating the spectrum of all degrees of freedom as continuous with {{} a continuous density of frequency levels} $\gamma_L(\Omega)$. With the continuity of $\gamma_L(\Omega)$ macroscopic irreversibility shows up in correlation functions. In addition, the continuity and smoothness assumption of the  {{} density of frequency levels}  is necessary in the formulation of equilibrium  thermodynamics. The very definition of entropy relies on these conditions (see e.g.~\cite{LL5}), $S(E)=\ln (\gamma_H(E) \cdot \Delta) $, where $\gamma_H(E)$ is the density of energy {{} levels} at energy $E$ and $\Delta$ is an arbitrary  fixed energy window sufficiently separated  from the lower value $1/t_e$ and the upper value $1/t_s$. Only then, temperature $T(E)$ can be identified by  $1/T(E):=\partial_E S(E)$. Energy as a continuous parameter of the macroscopic thermodynamic equilibrium state manifold could not be defined without the continuity condition of the density $\gamma_H(E)$. Thermodynamics with Legendre transform relations between thermodynamic potentials derived from partition sums of different  equilibrium ensembles relies on the smoothness condition as well and is justified up to relative fluctuations of order $1/\sqrt{N}$ in the thermodynamic potentials (see e.g.~\cite{Zia2009}).  Since thermodynamic entropy $S(E)$ is an extensive quantity,  the condition of continuity $t_s\ll t_e$ is easily justified as  $t_e\propto \gamma_L(E)\propto e^{S(E)}$ which is exponentially larger in $N$  than $t_s$. Since the  densities $\gamma_H(E)$ and $\gamma_L(\Omega)$ are related by convolution, $\gamma_L(\Omega)=(\gamma_H(-E)\ast \gamma_H(E))(\Omega)$, qualitative considerations of continuity and scaling with $N$ apply equally to both.

In addition to pure reversible mechanics, thermodynamics requires in its foundation an unavoidable statistical element for large numbers $N$: the assumption of continuously distributed energy eigenvalues due to an unavoidable resolution limit of the eigenvalues. This condition of continuous spectrum suffices to explicitly construct an irreversible equation of motion for relevant variables from the microscopic von Neumann equation valid under the condition $t_s\ll t_e$. This will be shown in Sec.~\ref{insta}.  It is a consequent combination of projection techniques pioneered  by  Nakajima \cite{Nakajima1959} and Zwanzig \cite{Zwanzig1960} and the resolvent method in the frequency picture pioneered by Fano \cite{Fano}.  The relevance of variables is conditioned by the fact that their spectrum under isolated conditions  could be resolved on the time scale $t_s$. By the projection to relevant variables the dynamics is no longer local in time. Retardation effects for relevant variables show up via integrating out irrelevant variables and the resulting dynamics is non-Markovian. It is therefore advantageous to change to the frequency picture of dynamic equations (resolvent equations) via the Laplace transform. In the following $f(\Omega+\imath \epsilon)$ denotes the Laplace transform of time dependent function $f(t)$, $f(\Omega+\imath \epsilon )=\int_0^\infty dt f(t) e^{i\Omega t-\epsilon t}$. Here $\Omega$ and $\epsilon$ are real valued frequency and positive  regularizing inverse time scale, respectively. 
For the von Neumann equation of all degrees of freedom this means 
\be\label{Int0}
\rho_{\rm tot} (\Omega+\imath \epsilon)=iG_{\rm tot}(\Omega+\imath\epsilon)\rho_{{\rm tot}_0} \, ,
\ee  where $\rho_{{\rm tot}_0}$ is an initial state and 
\be\label{Int1}
G_{\rm tot}(z)=\LBK z - L_{\rm tot} \RBK^{-1}
\ee is the resolvent of Liouville $L_{\rm tot}$ with poles at the real valued eigenfrequencies $\Omega_n$ symmetrically distributed around $\Omega=0$. 
When the limit to a continuous spectrum of $L_{\rm tot}$ has been taken, the limit $\epsilon\to 0^+$ can usually be performed. For systems with a many body localization (see e.g.\cite{Abanin2019})  phase with respect to a certain basis the global spectrum may be dense, but the typical local  density of {{} energy levels}  vanishes for $N\to \infty$. Thus, typically $t_e \to 0$  and no equilibration can take place. Such many body localization phases will not be considered further in the course of this work.

The exact relaxator Liouville equation for relevant variable states under the condition of continuous total spectrum will be discussed in Section~\ref{insta}.  It has the same structure as \Eq{Int0}. However, the initial state and the  resolvent are restricted  to the relevant degrees of freedom and depend on frequency showing memory effects.  The initial state $\rho(\omega)_0$ describes a frequency dependent initial state for relevant variables incorporating initial correlations between all degrees of freedom. Furthermore, the resolvent is now 
\be\label{Int3} 
G(z)=\LBK z - L(\omega) \RBK^{-1}
\ee 
for a frequency dependent non-Hermitian  relaxator Liouville operator $L(\omega)$ acting  on relevant variable states 
\be\label{Int4}
\rho (\omega)=iG (\omega+\imath 0^+)\rho(\omega)_0 \, .
\ee  
Equation \Eq{Int4} is the general equation of motion for relevant degrees of freedom in many body systems under the condition $t_s\ll t_e$. 

In Section~\ref{liouv} we derive and discuss the general properties of relaxator Liouville dynamics thereby relating $L(\omega)$ to the Hamiltonian of relevant variables interacting with irrelevant variables.  
Although there are several approaches to non-Markovian irreversible dynamics (see e.g.~\cite{BreuEtal}-\cite{ZhangEtAl2012}) we believe that the relaxator Liouville approach is most general and covers them in a systematic and transparent way, well suited for phenomenological, axiomatic,  and constructive modelings.  In Section~\ref{kinet} kinetic equations (see e.g.~\cite{Bonitz2016}) for one-particle densities are constructed as a special application of Liouville relaxator dynamics where one-particle degrees of freedom are taken as the relevant ones. The corresponding relaxator determines the self-energy in the corresponding nonequilibrium Green function NEGF (see e.g.~\cite{VLeuwen2006},\cite{Datta}). By incorporating driving fields for currents in the system we derive linear response susceptibilities for relaxator Liouville dynamics that generalize Kubo's linear response theory \cite{Kubo1957} and discuss different sources of irreversibility in Sec.~\ref{linea}.  By a weak coupling approximation in Sec.~\ref{weakc} it is possible to express the relaxator Liouville in terms of isolated properties of relevant and irrelevant variables.  We investigate the Markov approximation to compare our  general approach  with the well known  semi-group approach after \cite{Lindblad} and \cite{Gorini} ( for overviews see e.g. \cite{BreuPet}-\cite{Lidar2020})  to irreversible systems in Sec.~\ref{marko}. Finally we give a summary in Sec.~\ref{summa}.
\section{Instability of Reversibility}\label{insta}
We consider an arbitrary isolated quantum system denoted as total system that allows to distinguish two groups of degrees of freedom, relevant and irrelevant, there numbers being $N_s$ and $N_e$, respectively. The total Hilbert space is the product of the Hilbert spaces of relevant and irrelevant variables. {{} The relevant variables  are tracked up to $t_s$ which can be identified as the typical inverse level spacing of an isolated system of only relevant variables.
It is now a defining condition for the distinction between relevant and irrelevant degrees of freedom that the total system as well as the idealization of an isolated system of irrelevant variables has a spectral resolution time of the same order, which we denoted as $t_e$. It has to be larger than $t_s$ by a factor which is exponential in $N_e/N_s \gg 1$. The total system has a huge number $N=N_e+N_s\approx N_e$ of degrees of freedom {{} almost all} of which are irrelevant.}
It has become popular to denote such systems of relevant variables in contact with a huge number of irrelevant variables as  {open system} (in the sense of \cite{BreuPet}). The relevant variables are denoted simply as system variables and the irrelevant ones as environment.
Note however, that the notion of open system here  only means a system of variables which is not isolated but interacts with the  environment. It is possible, that such open system represents a closed system in the thermodynamic sense, i.e. a system that even in equilibrium exchanges energy with its environment but no matter. Both sets of variables (system and environment)  may act within the same volume (e.g. in Brownian motion or in the scattering of electrons with phonons or with lattice imperfections and impurities) and may be even based on the same microscopic particles (e.g. in the relaxation to stationarity  of a  fluid of particles by self-interaction). It is not necessary that the system exchanges particles with the environment, but it is also not excluded. The belonging of chosen degrees of freedom to system or environment however is held fixed in the modeling. In case  the environment  surrounds the volume of the system we call it a reservoir (e.g. a small piece of matter subject to radiation). In case that the reservoir is in thermal equilibrium, we call it a heat bath. 

The  system's  state is defined by virtue of a projector $\DP$ on the Hilbert-Schmidt space of linear operators over the Hilbert space of the total system,
\be\label{Ins1}
\rho_\DP:= \DP \rho_{\rm tot} = \LK\Tr_{\rm env} \rho_{\rm tot}\RK \otimes \rho_{\rm env}=: \rho\otimes  \rho_{\rm env}\, , 
\ee
where $\rho_{\rm env} $ is a fixed density operator of the environment and $\rho$ is the searched for density operator of the system. By tracing out the environment the density operator $\rho$ is of much lower dimension as $\rho_{\rm tot}$, but suffices 
to calculate the expectation value of any observable $A$ of the system  in state $\rho_{\rm tot}$,
\be\label{Ins2}
\Tr_{\rm tot} \rho_{\rm tot} A\otimes 1_{\rm env}=\Tr \rho A \, .
\ee 
We will use the following terminology. The  Hilbert-Schmidt space of linear operators over the Hilbert space of the total system is known as the total Liouville space of system plus environment. The projected subspace by $\DP$  is  the open system's Liouville space, which we also denote as $\DP$-space, the complement space to the open system's Liouville space  as $\DQ$-space. 
The  density operator $\rho$ is an element of the  $\DP$-space and can  be represented by a $d\times d$ dimensional density matrix when the idealized isolated system's Hilbert space is $d$-dimensional.
The complementary projector is denoted as $\DQ:=1-\DP$. Projector property $\DP^2=\DP$, $\DQ^2=\DQ$ and the complement property $\DP\DQ=\DQ\DP=0$ are fulfilled for any choice of the fixed normalized environmental density operator $\rho_{\rm env}$ in \Eq{Ins1}. Note that $\rho$ is independent of the choice of $\rho_{\rm env}$ as long as one proceeds without further approximations involving a specific choice of $\rho_{\rm env}$. 

One choice for $\rho_{\rm env}$ in an exact approach is singled out by requiring the projectors $\DP,\DQ$ to be Hermitian with respect to the Hilbert-Schmidt metric, 
\bea\label{1HS}
\LK \sigma_{\rm tot} | \rho_{\rm tot} \RK&:=& \Tr_{\rm tot} \sigma^\dagger_{\rm tot} \rho_{\rm tot} = \LK   \rho_{\rm tot} | \sigma_{\rm tot} \RK^\ast \label{1HS1}\, ,\\
\LK \rho_{\rm tot} | \rho_{\rm tot} \RK &\geq 0&\, ,\label{1HS2}\\
\LK \sigma_{\rm tot} | \DP \rho_{\rm tot} \RK^\ast&=&\LK \rho_{\rm tot} | \DP^\dagger \sigma_{\rm tot} \RK \, .\label{1HS3}
\eea
By writing the density operators in an expansion of system and environmental operators, $\rho_{\rm tot}=\sum_k \rho^{(k)}\otimes \rho_{\rm env}^{(k)}$,  one can conclude: the Hermiticity $\DP=\DP^\dagger$   is valid on the total Liouville space only if $\rho_{\rm env}=d_{\rm env}^{-1} 1_{\rm env}$, where $d_{\rm env}$ is the dimension of the environmental Hilbert space, while the Hermiticity of $\DP$ on the $\DP$-space is valid for any choice of $\rho_{\rm env}$. Hermiticity of $\DP$  is passed on to $\DQ=1-\DP$.
Therefore, we prefer the choice $\rho_{\rm env}=d_{\rm env}^{-1} 1_{\rm env}$ as long as we perform exact calculations on $\rho$ involving the total Liouville space. In Section~\ref{weakc} we will introduce a weak coupling approximation. Within this weak coupling approximation it is more appropriate to consider the environment to be already in a self-consistent stationary state $\rho_{\rm env}$ which, under equilibrium conditions, acts like  a heat bath on the system. Whenever one takes the choice of $\rho_{\rm env}$ not being proportional to $1_{\rm env}$ one cannot rely on the Hermiticity of $\DP,\DQ$ on the total Liouville space.

\begin{figure}[t]
\centering
\includegraphics[width=14cm]{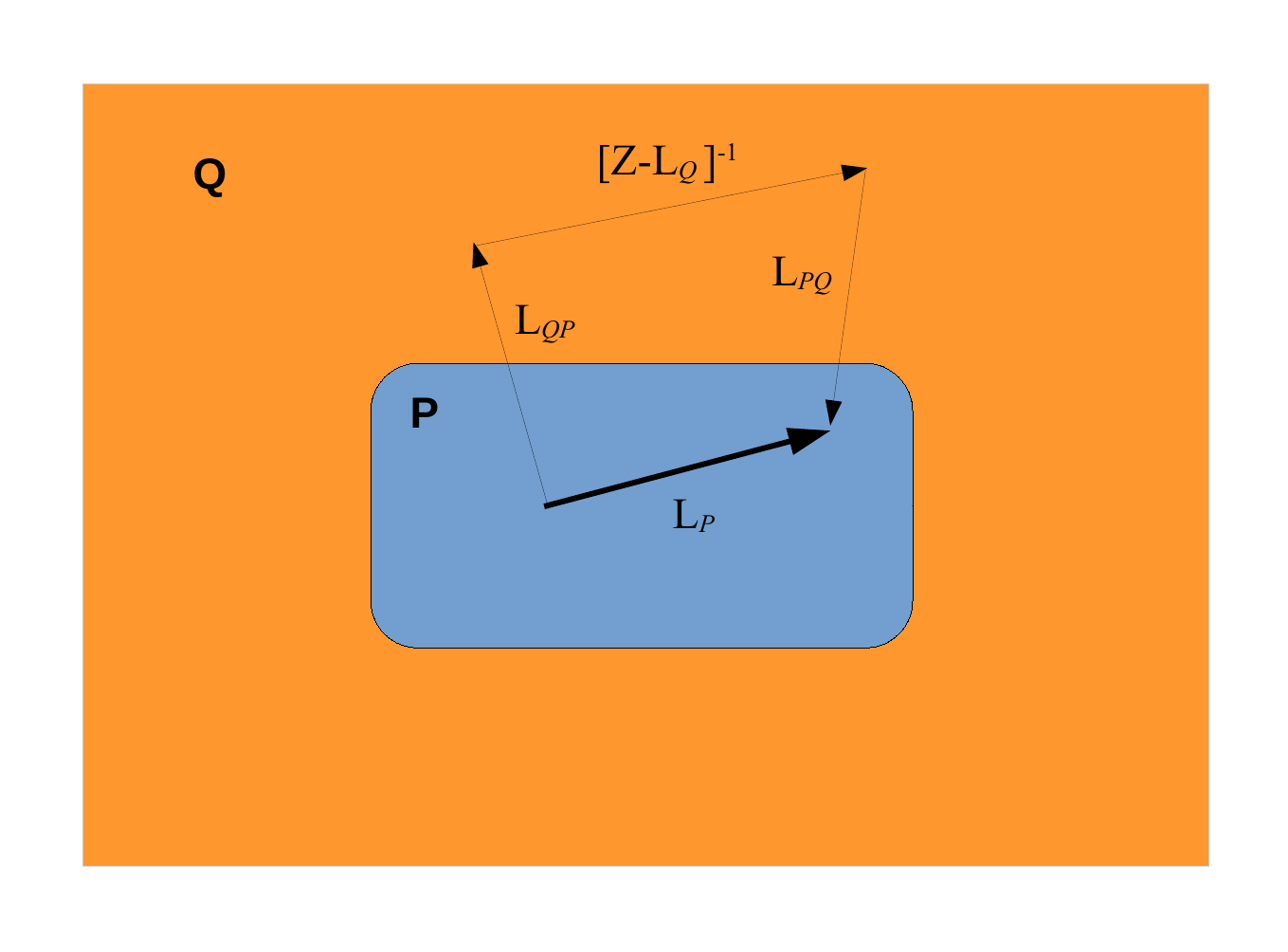}
\caption{Visualizing  relaxator Liouville dynamics in $\DP$-space by direct processes and virtual processes via hopping and propagation in $\DQ$-space.}
\label{FIG1}
\end{figure}

The total density operator will be decomposed as
\be
 \rho_{\rm tot} =\rho_\DP+ {\Delta\rho}^{\rm corr} \, , \; {\Delta\rho}^{\rm corr}:=\DQ\rho_{\rm tot} \, ,\label{4.51}
\ee
where ${\Delta\rho}^{\rm corr}$ captures correlations between system and environment. 
With  the decomposition of the total Liouville
\be
	\DL_{\rm tot} = \DL_\DP + \DL_{\DP\DQ} + \DL_{\DQ\DP} +\DL_\DQ \, , \label{44.Li}
\ee
 the equation of motion \Eq{Int4} follows by algebraic means \cite{Jan2018} and no approximation is involved. In the derivation it is essential that 
a projected total resolvent $\DP G_{\rm tot} (z)\DP$
 can always be expressed by the resolvent of a  Liouville operating solely on the Liouville space of the open system  ($\DP$-space),
\be
\DP G_{\rm tot} (z)\DP= G(z)=\LBK z-L(z)\RBK^{-1} \, , \label{444.66}
\ee
and the relaxator Liouville is algebraically given as
\be	
\DL(z) =  \DL_{\DP} + \DL_{\DP\DQ} G_\DQ (z) \DL_{\DQ\DP} \, . \label{4.57}
\ee
It has  the following intuitive interpretation: the first term $\DL_{\DP}$ is an isolated system Liouville in the absence of an environment {{} (it may however contain mean-field influences of the environment, see \Eq{33.7a} below)} whereas the second term
describes virtual processes in the system triggered by the environment, $\DL_{\DP\DQ} G_\DQ (z) \DL_{\DQ\DP}$, hopping to $\DQ$-space, taking there  a lift with isolated $\DQ$-propagator and finally hopping back to $\DP$-space, see  Fig.~\ref{FIG1}. This second term  is no longer Hermitian and breaks reversibility. 
\begin{figure}[t]
\centering
\includegraphics[width=14cm]{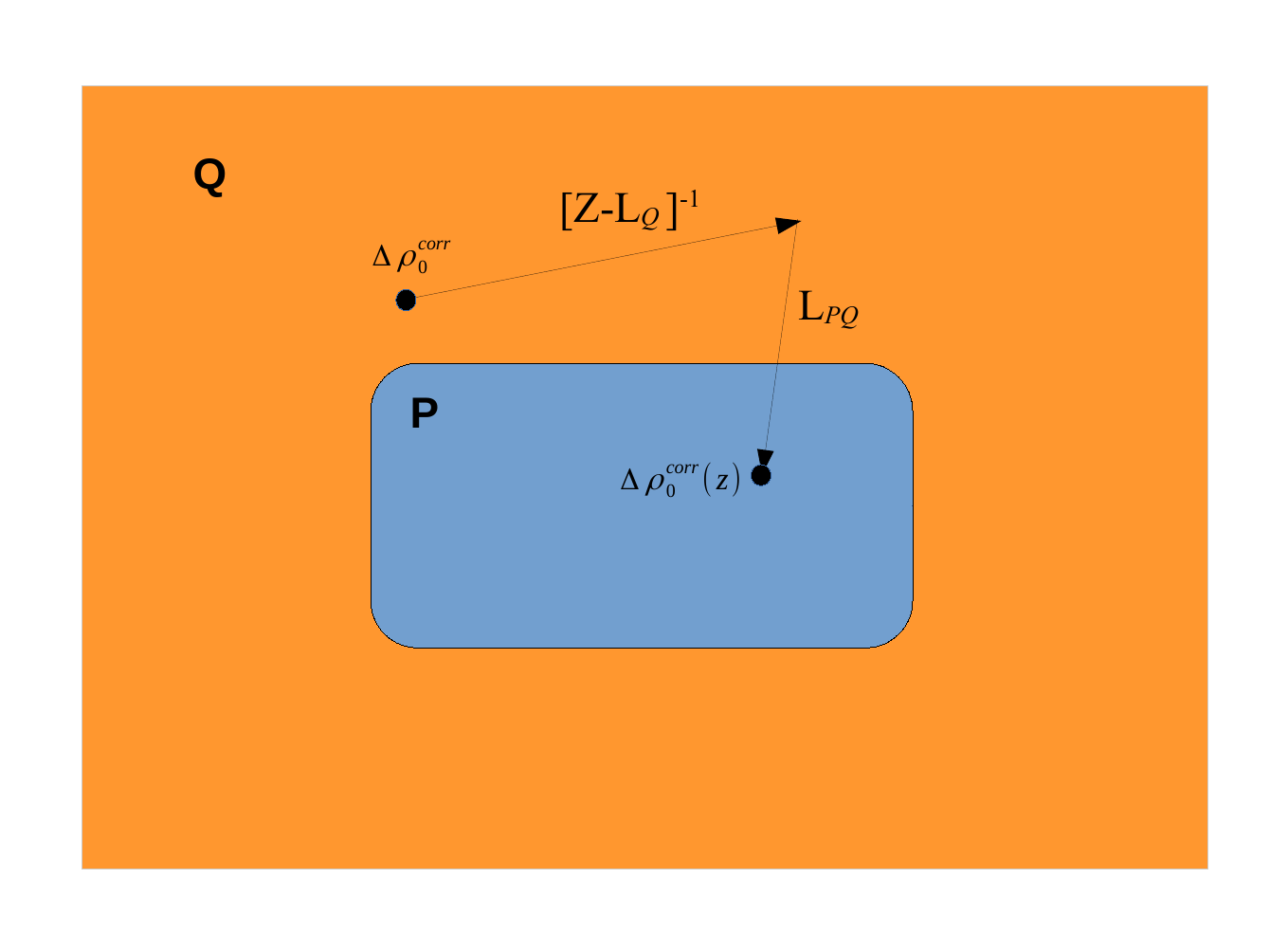}
\caption{Visualizing the contribution of virtual processes by initial correlations to the opens sytems's initial state. The correlation part $ {\Delta \rho}^{\rm corr}_0 $ of the initial state within the complementary $\DQ$-space  propagates there  and  hops to the $\DP$-space to contribute to the system's initial state in $\DP$-space with ${\Delta \rho}^{\rm corr}_0(z)$.}
\label{FIG2}
\end{figure}

The initial state $\rho_0(z)$ in \Eq{Int4}
is the initial state $\rho_0$ plus a virtual change of initial state within the system, caused by the initial correlation (${\Delta \rho}^{\rm corr}_0$) that gets a lift by the isolated $\DQ$-propagator and hops to $\DP$-space \cite{Jan2018} (see Fig.~\ref{FIG2}),
\be
	{\Delta\rho}^{\rm corr}_0(z)=  \DL_{\DP\DQ} G_\DQ(z) {\Delta \rho}^{\rm corr}_0 \, .\label{444.rho3} 
\ee
Equation (\ref{Int4}) is an exact equation of motion for states $\rho$ of the system, provided one knows $G_\DQ (z)$. 
$G_\DQ(z)$ can be considered as a propagator of an isolated system with Hermitian Liouville of the form  $\DL_\DQ \cdot = \LBK H_\DQ, \DQ\cdot\RBK$ with Hermitian $H_\DQ$ and corresponding real valued spectrum $E_a$.  If Hermiticity of $\DQ$ itself  is not guaranteed we can replace $L_\DQ$ with the total Liouville  $L_{\rm tot}$ since in \Eq{4.57} and in \Eq{444.rho3} the resolvent $G_\DQ (z)$ is framed by two $\DQ$-projectors. 

One may object that $L(z)$ relies on the solution of the total system, expressed by $G_\DQ (z)$. But we do not need the spectral fine structure of $G_\DQ (z)$. We work  with phenomenological and/or constructive modelings of $G_\DQ (z)$ as an input device to extract general properties and quantitative properties of the dynamics of system states without knowing a detailed solution of the total system. 

As to the general properties we decompose $G_\DQ (z)$ into Hermitian and anti-Hermitian parts, for $\epsilon \to 0+$,
\be
 G_\DQ(z)=\LBK z- \DL_\DQ \RBK^{-1} = {\rm P} \LBK \omega -\DL_\DQ \RBK^{-1}  -   \imath\pi \delta (\omega -\DL_\DQ) \, ,\label{4.57a}
\ee
where $ {\rm  P}$ stands for the Cauchy value on integration.
Due to the $\delta$ functional in \Eq{4.57a}, the Hermiticity of $\DL(z)$ is still valid for almost all values of $\omega$, except for a number of zero measure.
If, however, the spectrum $\Omega_n$ of complementary dynamics  cannot be resolved on the scale of the  time $t_{s} \ll t_e$  it has to be treated as distributed with a positive, even  {{} continuous density of frequency levels} $\gamma_{\small \DQ}(\omega)$ and we arrive at
\be
\DL(\omega) =  \DL_{\DP} +  {\cal D}(\omega) \, .\label{2.3}
\ee
where the presence of the  dissipator ${\cal D} (\omega)$
 shows the irreversible character of the dynamics described by $L(\omega)$.
The two contributions to the dissipator will be denoted as spectral shift $\Delta {\cal  H} (\omega)$ (due to the resemblance with the Lamb shift) and as ($-\imath$ times)  relaxator $\Gamma(\omega)$, 
\bea
 {\cal D}(\omega)&=& \Delta {\cal  H} (\omega) -\imath \Gamma(\omega)\, , \label{dissi2} \\ 
\Gamma(\omega) &=& \pi  \DL_{\DP\DQ} \delta(\omega- L_\DQ) \DL_{\DQ\DP} = \non\\ 
&=& -\lim_{\epsilon \to 0}  \frac{1}{2}
\DL_{\DP\DQ} \LK G_\DQ(z)-G_\DQ(z^\ast) \RK\DL_{\DQ\DP}  \, ,\label{dissi3}\\
  \Delta {\cal  H} (\omega) &=& \frac{1}{\pi}{\rm P} \int \d\Omega\, \frac{\Gamma(\Omega)} {\omega -\Omega} \label{dissi4}\, .
 \eea
 Equation \Eq{dissi4} shows  a Kramers-Kronig relation between the spectral shift and the relaxator.
With the convenient Hermitian choice of the projector $\DP$ the spectral shift is a Hermitian super operator and the relaxator is a Hermitian and positive super operator.

\begin{figure}[t]
\centering
\includegraphics[width=15cm]{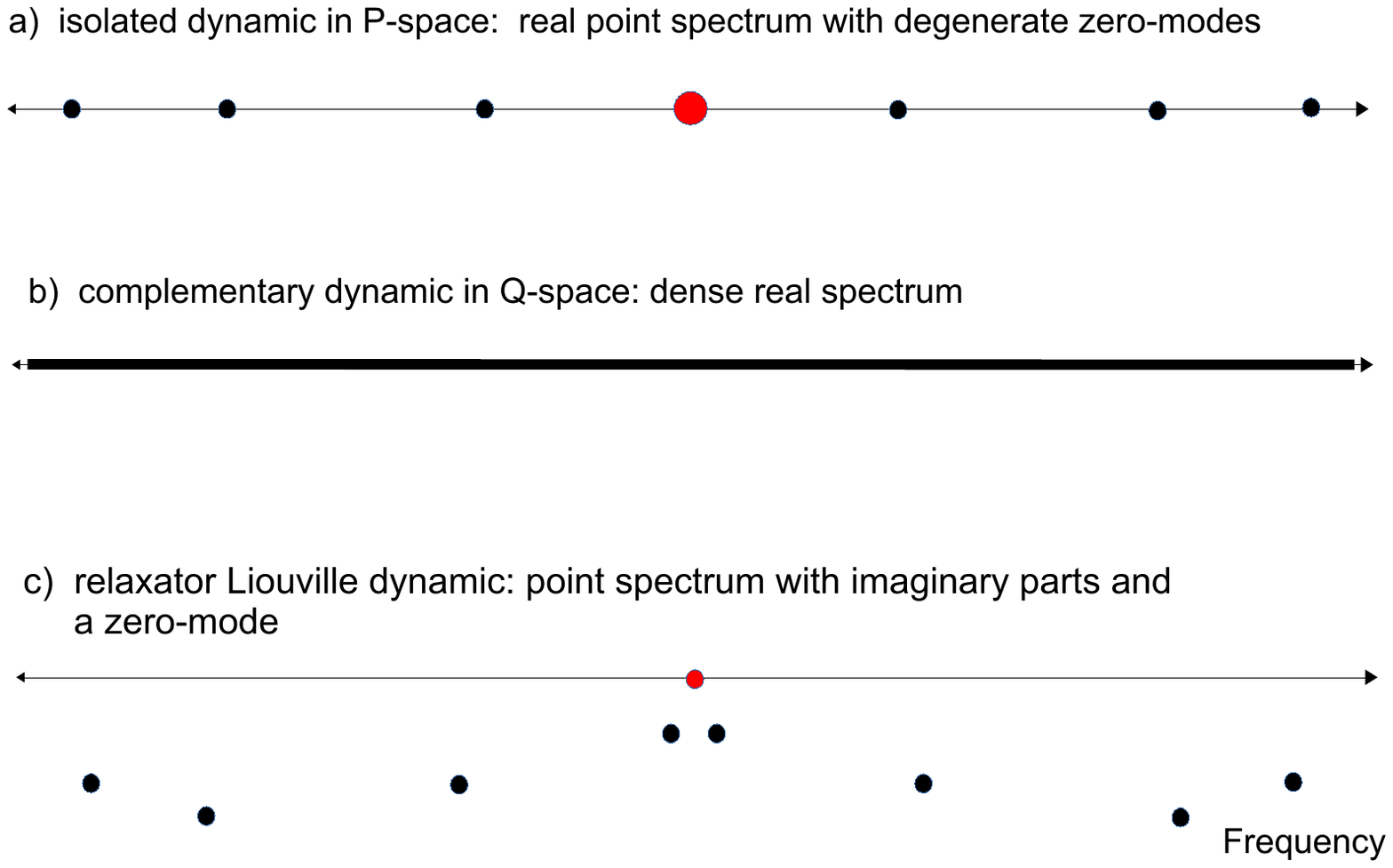}
\caption{Sketch of spectral properties of (a) the isolated system dynamic, (b) the complementary dynamic and (c) of the  relaxator Liouville dynamic.}
\label{FIG2c}
\end{figure}

 The presence of the relaxator is the central new element of the dynamics for states in a macroscopic system on time scales $t_s \ll t_e$. The relaxator is given by a resolvent $G_\DQ(z)$ within a continuous spectral resolution  and the hopping  super operators.
 It can also be given in terms of the  {{} continuous density of frequency levels}, $\gamma_{\small \DQ}(\omega)$,   corresponding eigenspace projectors $\Pi^\DQ(\omega)$, and the hopping super operators, $L_{\DP\DQ}$, $L_{\DQ\DP}$ as  
 \be
  \Gamma(\omega)=\pi  \gamma_{\small \DQ}(\omega)  L_{\DP\DQ} \Pi^\DQ(\omega) L_{\DQ\DP} \label{lioT}\, .
  \ee
So we can conclude as follows:
provided the system's energy spectrum can be considered as   discrete relative to the continuous spectrum of the complementary system,  the $\DQ$-propagator leads to  non-positive imaginary contributions to the eigenvalues of  the Liouville (see Fig.~\ref{FIG2c}), indicating that the system undergoes  a dynamic phase transition \cite{Pol2011} to irreversibility, with relaxation and decoherence (see Sec.~\ref{liouv}) when coupled to an environment \cite{JanBook2016}. A similar characterization of irreversibility by complex eigenvalues of the time-evolution operator within the unit-circle in many-body systems has been put forward in \cite{YoshiSa2025}.

It is worth mentioning that our criterion  $t_{s} \ll t_{e}$ for irreversibility was used before by van Kampen \cite{vKamp} to derive from reversible quantum mechanics a classical master equation for the probability to find a system in a macro state \cite{VKampInNoise}. Whereas van Kampen used  coarse grained modeling  over frequency levels of width $1/t_{s} \gg 1/t_{e}$ and brute force omission of fine structure in matrix elements of observables with respect to the total system's frequency spectrum to arrive at his goal,  we used an exact projector formalism to derive the equation of motion \Eq{Int4} for the system density operator.

\section{Relaxator Liouville dynamics: general properties}\label{liouv}
 Equation (\ref{4.57})  allows  to estimate the relevant time scale of relaxation indicating irreversibility. 
{{} The couplings $\DL_{\DP\DQ} \propto \DL_{\DQ\DP }$ can be made dimensionless by multiplication with an appropriate factor, $(\tau_{se})^{-1} L_{PQ}$. This factor  defines a time scale $\tau_{se}$ denoted as transition time for transitions between $\DP$- and $\DQ$-spaces. Since $\LBK z- \DL_\DQ \RBK^{-1}$ is the Laplace transform of the time evolution operator in $\DQ$-space the relaxator is the Fourier transform of a time dependent correlation function of coupling operators entering in $L_{\DP\DQ}$. Denoting the typical relaxation time of time dependent correlation functions of two environmental observables as environmental correlation time $\tau_e$, we can estimate the factor ${\tau_r (\omega)}^{-1}$ which makes $\Gamma(\omega)$ dimensionless  by dimensional analysis as 
\be\label{2.4}
\tau_r(\omega) = a \cdot \tau_{\rm se}\frac{\tau_{\rm se}}{\tau_{e}(\omega)}\, ,
\ee
with some dimensionless number $a$ of ${\cal O} (1)$ and $\tau_e(\omega)= \tau_e/(1+(\omega\tau_e)^2)$. Because $\Gamma(\omega)$ introduces non-positive imaginary contributions to the eigenvalues of  the Liouville, states can show relaxation to stationary situations (see also later in this section).  The system's relaxation time $\tau_r$  can be defined as the time scale after which the system’s state of relevant variables reaches its stationary value exponentially fast. As  an order of magnitude  estimation we can thus identify the system's relaxation time with $\tau_r(\omega\to 0) \approx \tau_{\rm se}({\tau_{\rm se}}/{\tau_{e}})$ by  \Eq{2.4}.

To discuss the importance of initial correlations expressed by  \Eq{444.rho3} we look at its contribution to the initial expectation value of any reasonable observable $A$ and ask when is it much smaller than the  expectation value in the state $\rho_0$,
\be\label{cormem1.2}
|\Tr \LB A  \DL_{\DP\DQ} \LBK z- \DL_\DQ \RBK^{-1} {\Delta \rho}^{\rm corr}_0 \RB | \ll |\Tr \LB A  \rho_0 \RB |\, .
\ee
We expect that the ratio of the two  sides of \Eq{cormem1.2} is set by  the typical time scales associated with   $\DL_{\DP\DQ}$ and with $\LBK z- \DL_\DQ \RBK^{-1}$, respectively. By similar dimensional analysis as before we find as a condition for neglecting initial correlations
\be\label{cormem1.3}
 \frac{\tau_{e}(\omega) }{\tau_{\rm se}} \ll 1 \, .
\ee
As a consequence, the estimation \Eq{2.4} of the  relaxation time
tells that $\tau_r(\omega)$  has to be much larger than the transition time, so that we have the inequality 
\be\label{cormem1.4}
\tau_{e}(\omega) \ll {\tau_{\rm se}}  \ll \tau_r(\omega)\, 
\ee
as the characteristic condition for the  initial correlations to be ignored.} If the restrictive condition \Eq{cormem1.4} is  violated, we have to face a significant influence of the initial correlations during the time evolution of the system until the stationary state is reached. This implies to  work with the equation of motion \Eq{Int4} to calculate quantities at intermediate times. A semi-group modeling ignores initial correlations and is not capable of the dynamics at intermediate times when  \Eq{cormem1.4} is  violated.  However, the stationary state $\rho_\infty$ is not influenced by initial correlations and completely determined by the zero mode of $L(z=\imath 0^+)$ (see later in this section). For the generic situation of a unique stationary state initial correlations are as unimportant as the initial state itself. 

The Hermiticity of density operators $\rho^\dagger(t)=\rho(t)$ leads for arbitrary choice of $\rho_{\rm env}$ for the projector in \Eq{Ins1} via the Laplace transform to the following properties of Hermiticity in the frequency domain:
\bea
\LK\rho(z)\RK^\dagger &=& \rho(-z^\ast) \, ,\label{hermi1}\\
\LBK L(\omega)\rho(\omega)\RBK^\dagger&=& -L(-\omega)\rho(-\omega)\, ,\label{hermi3}\\
\LBK {\rm H} (\omega) \rho(\omega)\RBK^\dagger&=&-{ \rm H} (-\omega) \rho(-\omega)\, ,\label{hermi4}\\
\LBK \Gamma (\omega) \rho(\omega)\RBK^\dagger&=&\Gamma (-\omega) \rho(-\omega)\, .\label{hermi5}
\eea
For the convenient choice of a Hermitian projector $\DP$ 
the spectral shift ${\rm H}(\omega)$ is a Hermitian super operator and the relaxator $\Gamma(\omega)$  is a positive  super operator and an even function of $\omega$. For our purposes this means  for any density operator $\rho$  
\be\label{gammasympos}
\Tr\rho\Gamma(\omega)\rho=\Tr\rho\Gamma(-\omega)\rho\geq 0\, .
\ee
With Hermitian projector $\DP$  the spectral shift vanishes at $\omega=0$,
\be\label{2.3c}
\Delta {\cal  H} (0) =0
\ee
due to the even character of the relaxator and the corresponding odd character of the spectral shift with respect to frequency $\omega$.

A modeling of an open quantum system can thus be based on the choice of an isolated system Liouville $L_\DP$ and  a choice of a positive, even relaxator $\Gamma(\omega)$,
which  has to fulfill the probability conservation condition when applied to any state $\rho$,
\bea
\Tr \LB L(\omega)\rho \RB &=&0 \, , \label{2.3d}\\
\Tr\LB \Gamma(\omega)\rho\RB &=&  -\imath\Tr \LB \Delta {\cal  H} (\omega) \rho \RB\, .\label{2.3dd}
\eea
Equations (\ref{2.3d},\ref{2.3dd}) express constraints for the relaxator Liouville dynamics.

The time dependent density operator is given by inverse Laplace transformation, 
\be\label{2.3e}
\rho(t) = \frac{\imath}{2\pi} \int_{-\infty + \imath0^+}^{\infty+\imath 0^+}  dz \, e^{-\imath zt}G(z) \rho_0(z)  \, . 
\ee
Although there is no  quantum dynamical map from the initial $\rho_0$ to $\rho(t)$ within the open system in a strict sense (the part $\Delta\rho^{\rm corr}_0(z)$   depends on initial properties of the environment),  the  generated dynamics \Eq{2.3e}  is as operative as the semi-group dynamics $\rho(t)=e^{-\imath \DL t}\rho_0$ to which it reduces when  initial correlations and memory effects can be neglected.

From the spectrum of eigenvalues of the  Liouville we can get the overall time behavior (see~\cite{Jan2017}).
To this end one takes the spectral representation of the relaxator Liouville for the resolvent
\bea
 \rho(z) &=& \imath \LK z -\DL(z) \RK^{-1}  \rho_0(z) \non\\
&=& -\imath \sum_{k=0}^N \frac{a_k(z)}{z-\lambda_k(z)}  {\sf R}_k(z) \, , z=\omega+i\epsilon \, , \epsilon\to 0^+\label{solution}\, .
\eea
Here ${\sf R}_k(z)$ is the right eigenoperator of the  relaxator Liouville corresponding to eigenvalue $\lambda_k(z)$ and 
\be\label{2.store}
a_k(z)= \LK  {\sf L}_k(z) \mid \rho_0(z) \RK
\ee
is the Hilbert-Schmidt scalar product between the corresponding left eigenoperator ${\sf L}_k(z)$ and the effective initial state $\rho(z)$. Left  and right eigenoperators are chosen bi-orthonormal (see~\cite{Jan2017}),
\be \label{bi-ortho}
\LK  {\sf L}_k(z) \mid {\sf R}_{k'}(z)\RK=\delta_{kk'}\, .
\ee

In the relaxator Liouville  the limit $\epsilon \to 0^+$ has been performed and $L(\omega)={\rm H}(\omega) -\imath \Gamma (\omega)$ has complex  eigenvalues $\lambda_k(\omega)$ and left eigenoperators ${\sf L}_k(\omega)$  and  right eigenoperators ${\sf R}_{k}(\omega)$.
Now for non-singular ${\sf R}_k(\omega), a_k(\omega)$  we define  the effective eigenvalues $z_k=\omega_k - \imath \delta_k$ with $\delta_k \geq 0$,  by 
\be\label{effeigen}
{\omega_k-\imath \delta_k= \lambda_k(\omega_k)} \, .
\ee
Then isolated poles $\rho(z)$ appear at those effective eigenvalues $\omega_k-\imath \delta_k$ (for non-vanishing $a_k(\omega_k)$)
and the inverse Laplace transform \Eq{2.3e} can be performed
with the method of residues yielding
\be\label{lastt}
           \rho(t) = \rho_\infty + \sum_{k=1}^N    \frac{1}{2} \LK  A_{k} {e^{-\imath\omega_kt}} +  A^{\dagger}_{k}{e^{\imath\omega_kt}} \RK e^{-\delta_k t}
	 \, . 
\ee
Here the stationary state $\rho_\infty$ has been separated and traceless right eigenoperators ${\sf R}_k(z_k)$ with factors $a_k(z_k)$ and  factors $1/(1-\frac{\d\lambda}{\d z}(z_k))$ have been put together to traceless operators $A_k$ and brought in a way making Hermiticity obvious. Hermiticity of $\rho(t)$ requires that effective eigenvalues occur in pairs, $z_k, -z^\ast_k$ for $\omega_k\not=0$.  
Taking the matrix elements in a basis where $\rho_\infty$ is diagonal,
\be
          \rho_{nm}(t) = p_n\delta_{nm} + \sum_{k=1}^N    \frac{1}{2} \LK  \LK A_{k}\RK_{nm} {e^{-\imath\omega_k t}} +  \LK A_{k}\RK^\ast_{mn}{e^{\imath\omega_k t}} \RK e^{-\delta_kt}
	 \,  \label{lastt2}
\ee
reveals that  the generic time behavior in a finite non-isolated quantum system is  a superposition of damped, maybe over-damped, oscillations. This is  in contrast to a finite isolated  quantum system which shows an almost-periodic time dependence, a superposition of oscillations without any damping. The relaxation time $\tau_r$ is given by the largest value of $1/\delta_k$ in case of well separated effective eigenvalues, or as a collective decay time in \Eq{lastt2} when the effective eigenvalues become dense with respect to observational resolution, as discussed in Sec.~\ref{intro} for correlation functions. It is worth mentioning that according to \Eq{lastt2} $\rho(t)$ stays positive semi-definite  if $\rho(0)$ is such density operator.

The most basic example for the relaxator Liouville dynamics is a modeling of a 2-dimensional Hilbert state space and, thus, a 4-dimensional Liouville space.  We choose  four effective eigenvalues in accordance with the constraints on the spectral properties: a zero eigenvalue of $L$ corresponding to  the stationary state $\rho_\infty$, an overdamped mode with negative imaginary eigenvalue $-i\delta_0$ and corresponding traceless right and left eigenoperators ${\sf R}_0$, ${\sf L}_0$, respectively. Finally we take  a pair of complex eigenvalues $\pm \omega_1 - i \delta_{1}$ ($\delta_1 >0$) with corresponding traceless right and left eigenoperators ${\sf R}_1$, ${\sf L}_1$, respectively. 
\be\label{2.44}
 (z-L(z))^{-1} \cdot = \frac{-i}{z} \Tr \LB {\cdot} \RB \rho_\infty  + \frac{{\sf R}_{0} \Tr \LB 
 {\sf L}_{0}^\dagger \cdot \RB }{z+ \imath \delta_0}
 + \frac{{\sf R}_{1} \Tr \LB 
 {\sf L}_{1}^\dagger \cdot \RB }{z-\omega_1 + \imath \delta_1} +  \frac{ {\sf L}_{1} \Tr \LB {\sf R}_{1}^\dagger \cdot  \RB}{z+\omega_1 + \imath \delta_1} \, .
\ee
In a frequency dependent signal $\Tr \LB A \rho(z)\RB $ of some observable $A$ we will find the stationary limit 
\be
A(t\to \infty)=  \lim_{z\to 0} -\imath z  \Tr \LB A \rho(z)\RB =\Tr \LB \rho_\infty A \RB
\ee
and resonances at oscillations with frequencies
\be 
\omega = \pm \omega_1\, ,
\ee
line width $\delta_1$ and coefficients $a=\Tr \LB A {\sf R}_1\RB \Tr\LB {\sf L}_1^\dagger \rho_0\RB$, $a^\ast $, respectively. The line width can be associated with a finite lifetime $\tau_1$ of oscillations with $\omega_1$,
\be
\tau_1=1/\delta_1, .
\ee
The smaller value $\delta_{\rm min}$ of $\delta_0,\delta_1$ dominates the relaxation of time signals $\Tr \LB A \rho(t)\RB$ to the stationary value with relaxation time $\tau_r=1/\delta_{\rm min}$. In the Appendix~\ref{specqubit} we treat explicitly  the example  of a qubit weakly coupled to a bath.

Now we turn to the relaxation to  a stationary state $\rho_\infty$ which cannot capture any information about the initial state ($a_0=1$ by probability conservation)  and its calculation without referring to  the full time dependence (see~\cite{Jan2018}).
States obeying the equation of motion \Eq{Int4} will generically  reach a unique stationary state in the long time limit. Due to the previously discussed time dependence of damped oscillations the long time limit can be taken as a long time average, defined by
\be
f_\infty=\lim_{\epsilon \to 0+} \epsilon \int_0^\infty dt\,  f(t) e^{-\epsilon t}  
\ee for a quite arbitrary time dependent function $f(t)$.    In the same sense it holds true that 
\be
	f_\infty=\lim_{z\to 0} -\imath zf(z) \, \label{3.1}\, ,
\ee
where $f(z)$ is the Laplace transform of of $f(t)$,  $f(z)=\int_0^\infty \d t \, e^{\imath zt} f(t)$.
By applying  this limit to the equation of motion \Eq{Int4} we find the following expression for the stationary state, 
\be
\rho_\infty = \lim_{z\to \imath 0^+} z \LBK z-\DL(z) \RBK^{-1}  \rho_0(z)   \, .\label{3.2}
\ee
Initial correlations are non-singular  for $z\to \imath 0^+$ in the initial state $\rho_0(z)$.
Then the long time limit is  determined by the  $z\to \imath 0^+$ limit of $z/(z-\DL(z))$ which is non-vanishing due to the existence of zero modes of $\DL(\imath 0^+)$.
Denoting the projector on the space of these zero modes by $\Pi^0_{\DL(\imath 0^+)}$,  the long time limit of the density operator reads
\be
	\rho_\infty = \Pi^0_{\DL (\imath 0^+)} \rho_0  (\imath 0^+) \, . \label{3.3}
\ee
The relaxator Liouville must have a zero mode due to probability conservation which means $\Tr \rho(z)=\imath /z$ and $\Tr \DL(z) \rho=0$ for every $z$ and $\rho$.
Thus, the relaxator Liouville has the unit operator $1$ as a  left eigenoperator with eigenvalue $0$ and, consequently, there must exist a right eigenoperator  with eigenvalue $0$.
In generic systems the zero mode will be non-degenerate  and leads to a unique normalized stationary state, $\rho_\infty$.
 The projector on the zero mode can be written as
\be
\Pi^0_{\DL(\imath 0^+)}= \rv{\rho_\infty}\lv{1}\, , \label{3.4}
\ee
where the notation implies the Hilbert-Schmidt scalar product.
Since the scalar product of the left unit matrix with the initial state amounts to taking the trace, any influence of the initial state is getting lost in a unique stationary state $\rho_\infty$. 
{{}  The relaxation of any initial state to a unique stationary state is at the heart of irreversible phenomenology and common experience. It has been studied within different modelings of irreversible systems classical and quantum, e.g. in \cite{vKamp}, \cite{BreuPet}, \cite{StefAlm} and \cite{MyöhEtAl}.}

In the non-generic case of  degenerate zero modes \Eq{3.4} will change to a sum of projectors on different eigenstate combinations of zero modes. It is obvious that in such case the scalar product of left eigenoperator with the initial state is not simply unity and these scalar products store information about the initial state. Stable degenerate zero modes must be  protected against perturbations by design or by some stabilizing natural mechanism. Only with stable degenerate zero modes initial conditions can influence the final stationary states. Such non-generic scenarios are discussed in \cite{Albert2018} for systems with semi-group dynamics. {{} Dependence of the long time behavior on initial states has also been observed within modelings with Luttinger liquids (see e.g. \cite{PerfEtAl}-\cite{BuchEtAl}). Luttinger liquids represent  fixed point models of one-dimensional interacting Fermions within unitary dynamics in the renormalization group sense. But they have an infinite number of conserved quantities which are however not stable against the breaking of unitarity by perturbations which are considered as irrelevant in the usual renormalization group sense of unitary dynamics. Thus, the dependence of the long time behavior on initial conditions is due to the degeneracy of the zero mode caused by constants of motion. The degeneracy and initial state dependence gets finally lost by small perturbations (see e.g.  \cite{MatAnd}-\cite{BuchEtAl}).}

As the unique zero mode of the relaxator Liouville dynamics at vanishing frequency, $\DL:=\DL(\omega=0)$,  the stationary state $\rho_\infty$ must satisfy the equation
\be
\DL \rho_\infty = \LBK H, \rho_\infty \RBK   - \imath \Gamma \rho_\infty = 0 \, , \label{statio}
\ee
where $\Gamma:=\Gamma(\omega=0)$ is the  relaxator at $\omega=0$ and $H$ the Hermitian Hamiltonian corresponding to $L_\DP$ . Due to \Eq{2.3c}  no spectral shift operator needs to be considered  in the equation for the stationary state for $\omega=0$.

As a condition  for the stationary state to be an equilibrium state, $\rho_{\rm eq}$, we require that both contributions in \Eq{statio} vanish separately, such that energy is conserved in the stationary state,
\be
\LBK H, \rho_{\rm eq}\RBK =0 \, . \label{energy}
\ee
In such equilibrium state $\rho_{\rm eq}$, characterized by \Eq{energy},  no global currents $I=\dot{X} =i \LBK H, X\RBK$ can go through the system on average, $\Tr I \rho_{\rm eq}=0$.
The energy conserved equilibrium state becomes unique in the relaxator Liouville dynamics by fulfilling  a second equation,
\be
\Gamma \rho_{\rm eq} =0\, . \label{master}
\ee
Thus, the stationary state $\rho_{\rm eq}$ is  diagonal in the energy representation $H\left. |n\KET = \epsilon_n \left. | n\KET$ and lies in the kernel of the  relaxator $\Gamma$.

To proceed in solving these equations we use the following representation of  super operators: an operator $X$ in any orthonormal basis $\LB \left. | n\KET\RB_{n = 1 \ldots d}$  with $\BRA n | m\KET=\delta_{mn}$ can  be written as 
$X = \sum_{nm} X_{nm} \left. | n \KET \BRA m |\right.$  with $X_{nm}=\BRA n |\right. X \left. | m\KET$.
Similarly, any  super operator $\cal A$ can then be written as
\be
{\cal A } \cdot = \sum_{mn,rs} {\cal A}_{rs,mn} \left. | r \KET \BRA s | \right. \BRA m | \right. \cdot \left. | n \KET \, , 
\ee
with 
\be
{\cal A}_{rs,mn} = \BRA r | \right. {\cal A} \LK \left.| m \KET \BRA n | \right. \RK \left. | s \KET \, .\label{comp1}
\ee
For $\Gamma$ the probability conservation requires $\Tr {\Gamma \cdot }=0$ leading to
\be
	\sum_{m} {\Gamma}_{mm,rs} =0 \; \; {\rm for} \;\; {\rm all} \; r,s\; .\label{cons1}
\ee
The stationary state is diagonal in the energy representation, $\rho_{\rm eq}=\sum_n p_n \left . | n \KET\BRA  n |\right. $ with non-negative probabilities $p_n$, and \Eq{master} reads
\be
0 = \sum_{n} \Gamma_{rs, nn} p_n \; \; {\rm for} \;\; {\rm all} \; r,s\; .
\ee
We specify to $r=s$ and decompose to insert probability conservation \Eq{cons1},
\be\label{pauli0}
0= \sum_{n_{\not= r}} \Gamma_{rr, nn} p_n + \Gamma_{rr,rr} p_r = \sum_{n_{\not= r}} \LK \Gamma_{rr, nn} p_n-\Gamma_{nn,rr}p_r\RK \, .
\ee
and arrive at a stationary Pauli master equation \cite{Pauli} with  $W_{rn}:=-\Gamma_{rr, nn}$ for $ n \not=r$ and $\sum_{n_{\not = r}} W_{nr} =-W_{rr}=\Gamma_{rr,rr} \geq 0$,
\be\label{pauli}
 \sum_{n_{\not= r}} \LK W_{rn} p_n- W_{nr}p_r \RK =0 \, .
\ee
$W_{rn}$ can indeed, for $n\not=r$, be expected to be non-negative transition rates by their relation to the positive relaxator, and particularly since  $\sum_{n_{\not = r}} W_{nr} =\Gamma_{rr,rr}\geq 0$. Non-negative transition rates $W_{rn}$ describe a gain of probability at $r$ by transitions from $n$.
A relabeling of states should not change this character. Since some of the off-diagonal $W_{nr}$ have to be non-negative we  expect that all of them are non-negative. It can explicitly be shown e.g. in the weak coupling approximation (see Appendix~\ref{robustenvmod}).

Equation (\ref{pauli}) is commonly known to describe the stationary equation of irreversible evolution into equilibrium states when $W_{rn}$ can be seen as transition rates for $r\not =n$ and when they fulfill the detailed balance condition,
\be
 W_{rn} p_n - W_{nr} p_r =0 \; \; {\rm for} \;\; r\not=n \, .\label{equip}
\ee 
The detailed balance condition expresses  micro reversibility of the dynamics: for the joint probability to find  states $r$ and $n$, separated by one time step, it does not matter which one of both was earlier. This is exactly expressed by \Eq{equip} when $W_{rn}$ are transition rates for one time step from state $n$ to state $r$. Since $W_{rn}=-\Gamma_{rr,nn}$ is build from the unitary  complementary dynamics it respects micro reversibility.
From \Eq{equip} we finally arrive at  the stationary distribution  by recursion,
\be\label{finaleq}
p_n=\frac{W_{n0}}{W_{0n}} p_0 \, .
\ee
We further distinguish two situations: (A) idealized isolated system with $d$ states in an energy shell of fixed energy  and (B) a system with an environment in equilibrium with a huge number $N$ of constituents.

In case (A) the transition rates within the energy shell are all equal. This is the case  according to
Fermi's Golden Rule as well as according to the principle of ignorance when the energy window $\Delta$ is limited by $1/t_e\ll \Delta \ll 1/t_s$. Therefore, one arrives at  the micro canonical
uniform distribution,
\be
p_n\equiv 1/d \, .
\ee

In case (B) the transition rate to the
energetically higher state is by the Boltzmann factor 
\be\label{caneq}
\frac{W_{nr}}{W_{rn}}=c e^{-(\epsilon_n-\epsilon_r)/T}
\ee
  smaller than for the reverse process. In \Eq{caneq}  $T$ is the temperature of the environment and $c$ a constant.
The condition \Eq{caneq} is expected to be fulfilled when $N$ becomes macroscopic. The argument is robust and independent of many details of $W_{nr}$.  The joint probability $p_rW_{nr}$ can be seen as a joint probability of system and environment traced out over the environment, $p_rW_{nr}=\sum_{\alpha\beta} p_{r\alpha}W_{n\beta;r\alpha}$ (see also \Eq{GammaInH} below). For macroscopic environment in equilibrium not influenced by the system we can assume  $p_{r\alpha}=p_rp_\alpha$ to factorize and $W_{n\beta;r\alpha}$ to be constant  transition rates within an energy shell at total energy $E_{\rm tot}=\epsilon_r+E_\alpha=\epsilon_n+E_\beta$. Since $p_\alpha$ can be related to the entropy $S(E_\alpha)$ as $p_\alpha \propto e^{-S(E_{\rm tot}-\epsilon_r)}$, to find the dependence on $\epsilon_r$ we expand $S$ to linear order as the dominant  one for large $N$ and find $W_{nr}\propto e^{\epsilon_r/T}$ with $T^{-1}:=\partial_E S(E)$ defining the environmental temperature.
The equilibrium distribution of the system then turns out to be the canonical distribution,
\be
p_n=\frac{e^{-\epsilon_n/T}}{\sum_{k=1}^d e^{-\epsilon_k/T}}\, ,
\ee
with  temperature $T$ inherited from the environment. In Appendix~\ref{robustenvmod} we show the property \Eq{caneq} explicitly within in the weak coupling approximation.

We close this section on general properties of the relaxator Liouville dynamics by presenting it in terms of a Hamiltonian modeling.
The total Hamiltonian $H_{\rm tot}$ consists of three parts, $H$ representing the isolated system,  $H_{\rm env}$ representing the isolated environment, and an interaction part, $H_{\rm int}$,
\be\label{33.1}
H_{\rm tot}=H + H_{\rm env} + H_{\rm int}\, .
\ee
To get  more details of the representation  we  take the most general form  of interaction,
\be\label{33.4}
H_{\rm int} = \sum_k S_k \otimes B_k\, ,
\ee where $S_k$ are Hermitian operators on the system's Hilbert space and $B_k$ are Hermitian operators on the environment's Hilbert space with expectation values $\BRA B_k\KET_{\rm env}:= \Tr_{\rm env} B_k \rho_{\rm env}$ and deviations from their expectation values $\Delta B_k := B_k - \BRA B_k\KET_{\rm env}$. 
We are then led to the following  expressions 
\bea
\DL_\DP \rho \otimes \rho_{\rm env} &=& \LBK H_\DP,\rho \RBK  \otimes \rho_{\rm env} \, , H_\DP:= H+  \sum_k \BRA B_k\KET_{\rm env} S_k \, ,\label{33.7a}\\
\DL_\DQ \DQ &=&  \LBK H_\DQ , \DQ \RBK \label{33.7b}:= \LBK H_{\rm env}+H + H_{\rm int} , \DQ \RBK -  \LK\Tr_{\rm env} \LBK H_{\rm int}, \DQ \RBK\RK \otimes \rho_{\rm env} \, ,\\
\DL_{\DQ \DP} \rho \otimes \rho_{\rm env} &=&\sum_k \LBK  S_k\otimes \Delta B_k ,  \rho \otimes \rho_{\rm env} \RBK\, ,\label{33.7c} \\
\DL_{\DP \DQ} \cdot &=&   \LK\Tr_{\rm env} \sum_k \LBK S_k\otimes \Delta B_k, \cdot \RBK \RK \otimes \rho_{\rm env}\label{33.7d} \, .
\eea
The  relaxator Liouville $L(\omega)$
reads 
\bea
 L(\omega) \rho  &=& \LBK H_\DP,\rho \RBK \nonumber + \\
&+& \Tr_{\rm env}  \LBK \sum_k S_k\otimes \Delta B_k , G_\DQ(z) \sum_{k'} \LBK S_{k'} \otimes \Delta B_{k'},\rho \otimes \rho_{\rm env} \RBK \RBK\, ,\label{33.8bliou}\, 
\eea
and the relaxator is
\be
\Gamma(\omega) \rho  =\Tr_{\rm env}  \LBK \sum_k S_k\otimes \Delta B_k , \pi\delta(z- L_\DQ) \sum_{k'} \LBK S_{k'} \otimes \Delta B_{k'},\rho \otimes \rho_{\rm env} \RBK \RBK\, .\label{GammaInH}
\ee
In these expressions  $\epsilon$ is send to $0+$ after taking the trace over the environment with a continuous spectrum.

 The  Hamiltonian of the complementary dynamics, $H_\DQ$,  has an eigenvalue representation, $H_\DQ \left. \mid m \alpha \KET =E_{m\alpha} \left. \mid m \alpha \KET $ where the factors  $\left. \mid m\KET$ belong to the system's Hilbert space and the factors $\left. \mid \alpha \KET$ to the environment's Hilbert space. 
The resolvent can be written in this eigenvalue representation and the  relaxator Liouville of \Eq{33.8bliou} reads in the $H_\DQ$ eigenvalue representation
\bea
L(\omega) \rho  &=&  \LBK H_\DP, \rho\RBK + {\cal D}(\omega)\rho \label{33.11b1}\, , \\  {\cal D}(\omega)\rho &=&  \sum_{\scriptsize {k,k',m,n,\alpha,\beta}}  \LB \frac{\LK \Delta B_k \RK_{\beta \alpha} \LK \Delta B_{k'}\rho_{\rm env} \RK_{\alpha \beta}}{\omega +\imath\epsilon - \LK E_{m \alpha}-E_{n \beta}\RK}  \LBK  S_k,  P_m S_{k'} \rho P_n\RBK   
\right. + \nonumber \\
&+& \left. \frac{\LK \Delta B_k \RK_{\beta \alpha} \LK \rho_{\rm env}\Delta B_{k'} \RK_{\alpha \beta}}{\omega +\imath\epsilon - \LK E_{m \alpha}-E_{n \beta}\RK}  \LBK  P_m \rho S_{k'} P_n, S_k \RBK \RB \, ,  \label{33.11b}
\eea
where $P_m:= \left.\mid m\KET \BRA  m \mid\right. $ and $\epsilon\to0^+$ after summing over $\alpha, \beta$  with a continuous spectrum of states.

With a representation of $\Delta\rho_0$ by a superposition of system times environment operators,
\be\label{inidecomp}
\Delta \rho_0=\sum_l \Delta \rho_{{\rm s}0}^{(l)}\otimes \Delta \rho_{{\rm e}0}^{(l)}\, ,
\ee
we can calculate the correlation contribution to the  initial state as
\be\label{inicorham2}
\Delta \rho_0(\omega)=\sum_{k,l,\alpha,m,\beta, n} \frac{(\Delta B_k)_{\beta\alpha}(\Delta \rho_{{\rm e}0}^{(l)})_{\alpha \beta} \LBK S_k, P_m \Delta \rho_{{\rm s}0}^{(l)}P_n \RBK} {\omega+\imath \epsilon-(E_{m\alpha } -E_{n\beta })}\, ,
\ee
where $\epsilon\to0^+$ after summing over $\alpha, \beta$  with a continuous spectrum of states.

Equation~(\ref{33.11b1}) with \Eq{33.11b} is a general representation of the relaxator Liouville in terms of a decomposed Hamiltonian for the total system, ready for further investigations. For  a frequency resolution of order $1/t_{s}\gg 1/t_{e}$ we  may factorize the contributions in \Eq{33.11b} to ${\cal D}(\omega)$ coming from the environment and the system, respectively. Such factorization of the form
\be\Gamma(\omega) \rho =  \sum_{\tilde{\omega}} \sum_{kk'} {\gamma}_{kk'}(\omega+\tilde{\omega}) \LBK  S_k,  (S_{k'} \rho)(\tilde\omega)  \RBK  +  h.c.   \label{factor1}\ee 
with an environmental correlation function $\gamma_{kk'}(\Omega)$ 
becomes  possible if  we can rewrite the total energies $E_{\alpha m}$ as additive with respect to system and environment, $E_{m\alpha }\approx E_\alpha + \epsilon_m$ and treat interaction energies between system and environment as small compared to the resolution scale $1/t_s$. This will be undertaken in Sec.~\ref{weakc}.

\section{Kinetic equations from relaxator Liouville dynamics}\label{kinet}
So far, the relaxator Liouville equation \Eq{Int4} was derived for an arbitrary open system.  To extract the  full time dependence of a many particle state, $\rho=\rho_{_{1\ldots N}}$, is not only impossible in practice, but also often not of interest; it contains to much information to be manageable. Therefore, one prefers to have an equation for the time evolution of the reduced density operator for just one particle or a few, say $s$, particles out of the $N$. In particular, the one-particle density operator suffices to calculate expectation values of additive quantities like kinetic energy and current.

Dynamic equations for one-particle distributions or density operators are called kinetic equations. Historically the Boltzmann equation \cite{Boltzmann1872} was the first of these and with it began the study of irreversible processes, including the relaxation to equilibrium. The Boltzmann equation and more general quantum kinetic equations can be derived  from the  BBGKY hierarchy (derived by Bogoliubov, Born, Green, Kirkwood, Yvon and others in the 1930's to 1940's)  for $s$-particle reduced density operators (for an overview  see \cite{Bonitz2016}) after appropriate decoupling approximations. 
We take the following definition for the $s$-particle reduced density operator $\rho_{_{1\ldots s}}$ 
\be\label{kin1}
\rho_{_{1\ldots s}} := \Tr_{_{{s+1}\ldots N}} \rho_{_{1\ldots N}} \; ; \; \Tr_{_{1\ldots s}} \rho_{_{1\ldots s}}= 1\, .
\ee
Other definitions in the literature differ by normalization conventions.

An alternative way to the BBGKY route to arrive at kinetic equations uses so called nonequilibrium Green functions (NEGF) (invented by Schwinger, Feynman and Vernon, Kadanoff and Baym, Keldysh \cite{Neqinventors} and others in the 1960's). These are nonequilibrium expectation values for one quantum object (particle) to propagate from one event $1$ to another event $2$. In the NEGF approach, the idea of a self-energy $\Sigma$ is exploited. It takes the many particle influences on the one-particle propagator into account on its way from $1$ to $2$. For an introduction to NEGF see \cite{VLeuwen2006}, Chap.~13 in \cite{Bonitz2016} and \cite{Datta}. 
Usually, in the NEGF approach, the self-energy is constructed by approximate methods to incorporate as much of the interaction between many particles as possible or needed. The self-energy can then show finite broadened levels shifted from stationary one-particle levels. {{} An up to date linearly in time numerical method for solving closed equations of one-particle density matrices in the framework of NEGF in the time domain for coupled electron-boson dynamics is presented in  \cite{KarlEtAl}. It satisfies conservation laws despite being perturbative. }

The route to kinetic equations we take with relaxator Liouville dynamics is close in spirit to the NEGF method, but the construction of the self-energy is formally exact {{} and non-perturbative} owing to a  projector method. Our procedure has the advantage to immediately leading us to  a closed equation for the one-particle density operator via a one-particle propagator.  It makes the origin of irreversible behavior apparent by the relaxator appearing in this equation. The corresponding spectral shift and relaxator in our approach correspond to the self-energy in the NEGF approach. 

We take a projector onto one-partice states in the following form
\be\label{kin4}
\PE \rho_{_{1\ldots N}} = \rho_1 \otimes \rho^0_{_{2\ldots N}} \, .
\ee
where $\rho_1$ is constructed from tracing out all particles but one, as defined in Eq.~\Eq{kin1}. 
The $N-1$-particle operator $\rho^0_{_{2\ldots N}}$, with $\Tr_{_{2\ldots N}}\rho^0_{_{2\ldots N}}=1$,  is taken as a fixed factor, independent of the chosen $\rho_{_{1\ldots N}}$. By omission of this fixed factor in the end, a reduced equation of motion on a one-particle Liouville space will result. 

It is easy to show that an application of $\PE$ to $\PE \rho_{_{1\ldots N}}$ does not change the projection. This establishes $\PE={\PE}^2$ as a projector. The complement projector is defined by
\be\label{kin5}
\QE=1-\PE
\ee
and fulfills the desired projector and complement rules,
\be\label{kin6}
{\QE}^2=1-\PE=\QE \; ;\; \PE\QE=\QE\PE=0 \, .
\ee
By the same algebraic reasoning as used in \cite{Jan2018} one can then show that
for a given $N$-particle Liouville $L_{_{1\ldots N}}$  we have
\be\label{kin7}
\PE  (z-L_{_{1\ldots N}})^{-1} \PE = (z- L_1(z))^{-1} \, 
\ee
which acts, like $L_1(z)$, only on the one-particle Liouville space ($\PE$-space).
The one-particle relaxator Liouville dynamics is then defined with the help of $L_1(z)$,
\be\label{kin8}
\rho_1(z)= (z- L_1(z))^{-1} \rho_{1;0}(z)\, .
\ee
The expressions for $L_1(z)$ and the initial state $\rho_{1;0}(z)$, containing initial many-particle correlations, take the form 
\bea
L_1(z) &=& L_1 + L_{\PE\QE } \LBK z - L_{\QE} \RBK^{-1} L_{\QE\PE }   \label{kin9}  \, ,  \\
\rho_{1;0} (z) &=& \rho_1 (t=0) + L_{\PE\QE } \LBK z - L_{\QE } \RBK^{-1}  \rho_{_{1\ldots N}}(t=0)  \label{kin10} \, ,
\eea 
with $L_1 : =\PE L_{_{1\ldots N}}  \PE$, $L_{\PE\QE }  := \PE L_{_{1\ldots N}} \QE$, 
$L_{\QE\PE } := \QE L_{_{1\ldots N}} \PE$, 
and $L_{\QE } := \QE L_{_{1\ldots N}}$.
The time dependent one-particle density operator is given by the inverse Laplace  transformation,
\be\label{kin2.3e}
\rho_{1} (t)  = \frac{\imath}{2\pi} \int_{-\infty + \imath0^+}^{\infty+\imath 0^+}  dz \, e^{-\imath zt} \LBK z - L_1(z) \RBK^{-1}\rho_{1;0} (z)  \, . 
\ee

We can  identify the source for relaxation of the one-particle density operator as the interaction with the other $N-1$ particles: as soon as  the spectrum of the complementary dynamics (described by ${L}_{\QE }$) becomes dense relative to the spectrum of the isolated one-particle dynamics (described by $L_1$),  it is a source of irreversibility and a relaxator can be defined for the kinetic equation (by the same reasoning that led to Eq.~\Eq{2.3}) by 
\be\label{kin21}
\Gamma_1 (\omega)=  {L}_{\PE\QE} \delta \LK\omega -L_{\QE}\RK {L}_{\QE\PE}\, . 
\ee

To set up the kinetic equation for one particle, one has to find  the propagator $(z-L_{\QE })^{-1}$. 
Although $ L_{\QE} $  defines  an interacting many-body problem as complicated as the original $L_{_{1\ldots N}}$ problem,
 for the kinetic equation we don't need the spectral details of $L_{\QE}$ for investigation times up to $t_s$. We can therefore introduce approximations, e.g. in a first step treating $L_{_{2\ldots N}}$ as an effective additive single-particle Liouville,  and still reach the goal for a kinetic equation for particle 1, capturing irreversible  relaxation to a stationary state of \Eq{kin9}.

\section{Linear response for relaxator Liouville dynamics}\label{linea}
Linear response theory addresses  deviations in expectation values of some  observable $B$ to linear order in a given external, in general time dependent, driving field $F(t)$, coupled to the system by some observable $A$. 
It is assumed that the  field causes a linear shift of the system's energy,
\be\label{LR1.0}
H \to H(t)=H - A\cdot F(t)\, .
\ee
For a time translational invariant and  causal linear response the resulting expectation value of $B$ can be written as
\be\label{LR2.0}
\BRA B\KET_t = \BRA B \KET_{\rm st} + \int_{-\infty}^{t} dt' \, \chi_{BA}(t-t') F(t') \, ,
\ee
where $\BRA B \KET_{\rm st} $ is the stationary expectation value of $B$ in the absence of the field and $\chi_{BA}(t'-t)$ is the phenomenological definition of the dynamical susceptibility, the  quantity of interest in linear response theory. Causality requires that $\chi_{BA}(t-t') =0$ for $t < t'$. 

The linear response deviation $\Delta B(t):=\LK \BRA B\KET_t - \BRA B \KET_{\rm st}\RK$ and the field $F(t)$ can be analyzed with respect to their frequency content.
By the convolution property of the  Laplace transform  and the condition of causality  one concludes the linear response relation in frequencies $z=\omega+\imath \epsilon\, , \epsilon \to 0^+$,
\be\label{LR4.0}
\Delta B (z) = \chi_{BA}(z) F(z) \, , 
\ee
where the functions with argument $z$ stand for the Laplace transform of the functions with time argument, e.g. 
$\chi_{BA}(z)=\int_0^\infty  d(t-t') \chi_{BA}(t-t') e^{\imath z(t-t')}$ with $z=\omega+\imath \epsilon$ for $\epsilon \to 0^+$.
The introduction of $\epsilon >0$ and the use of the Laplace transform  is mathematically adopted to the causal structure of the response: fields $F(t')$  influence the signal at time $t$ only when $t'\leq t$. 

Before turning to set up a linear response theory for open systems  we  recall the dynamical susceptibility formula found  by Kubo  \cite{Kubo1957}  to make a comparison easier. The Kubo formula is derived  for  thermally isolated Hamiltonian systems subject to weak external driving fields. Kubo showed in his famous paper that dynamic susceptibilities can be expressed  in terms of correlation functions in the stationary state in the absence of the driving field. The Kubo formula can be viewed as a generalization of the previously known fluctuation-dissipation theorem (see his review article \cite{Kubo1966}). 
It reads 
\be\label{K1.14}
\chi_{BA}(z) = -\imath \int_{0}^\infty \d s \,  e^{\imath z s}  \Tr\LB  \rho_{\rm eq} \LBK A, B(s) \RBK \RB \, , z=\omega +\imath \epsilon\, , \epsilon \to 0^+ \, ,
\ee
for the dynamic susceptibility of an isolated system, where $B(s) = e^{\imath H s} B e^{-\imath H s}=e^{\imath L s} B$.
With 
\be\label{K1.15}
\Tr\LB  \rho_{\rm eq} \LBK A, B(s) \RBK \RB = -\Tr\LB   \LK \LBK e^{-\imath L s} \delta L \RBK  \rho_{\rm eq}\RK B \RB
\ee
 and  carrying out the time integral sets \Eq{K1.14} in the  compact resolvent form, 
\be\label{K1.16}
\chi_{BA}(z) = -  \Tr\LB    \LK \LBK G(z) \delta L \RBK \rho_{\rm eq}\RK  B\RB \, , \epsilon \to 0^+\, .
\ee
Here $G(z)=[z-L]^{-1}$ and $\delta L\cdot :=[A,\cdot]$.
With Equation~\Eq{K1.16} it is easy to see that $\chi_{BA}(z)$ is analytic in the upper $z$-plane and has singularities at the spectrum of $L$. The limit $\epsilon \to 0^+$ thus requires a careful analysis of the spectrum of $L$.
Typically, a limit to an infinite number of variables or parameters has to be performed first to reach a reasonable limit $\chi_{BA}(\omega + \imath 0^+)$.

The Kubo formula has proved to be very successful. Irreversibility indeed shows up in dynamic susceptibilities like  dissipative conductances. However, when performing  measurements of dynamic susceptibilities,  the initial state is not necessarily an equilibrium state, and the number of variables is  huge, but finite. In the following we will alternatively start from an open quantum system, where we have gained insight how irreversibility  and stationarity emerges. It is a systematic approach to linear response in open systems based on the general relaxator Liouville dynamics and more general than other approaches \cite{BanEtAl2017}-\cite{AvronEtAl2012} to open systems that rely on semi-group or time local quantum master equations.  In \cite{WeiYan2011} an alternative general approach to linear response in open systems is presented. It is based on the Feynman-Vernon influence functional \cite{FeynVern} and hierarchical equations of motion  for propagator and state, which have to be solved. We prefer the relaxator  Liouville approach for which we know already general properties of the propagator and state as outlined in Sec.~\ref{liouv}.

We consider an open system consisting of system and its environment as in Sec.~\ref{insta}. The total Liouville is  time dependent by coupling to an external driving field,
\be\label{HJ1.0}
L_{\rm tot} (t)=L_{\rm tot} -\delta L  {F}(t)\, .
\ee
Note, $\delta L$ leaves the environment untouched.
The Laplace transform of the von Neumann total equation of motion  ${\dot \rho}_{\rm tot}(t)=-\imath L_{\rm tot}(t) \rho_{\rm tot} (t)$ reads
\be\label{HJ1.2}
-\rho_{\rm tot} (t=0)-\imath z \rho_{\rm tot} (z) = -\imath L_{\rm tot} \rho_{\rm tot} (z) + \imath  \sum_{\tilde \omega} \delta L \rho_{\rm tot}  (z-\tilde{\omega})  {F}_{\tilde{\omega}}\, , 
\ee
where the field $F(t)$ is given by its spectral representation
\be\label{HJ1.2a}
F(t)=\sum_{\tilde{\omega}} F_{\tilde{\omega}} e^{-\imath \tilde{\omega}t} \, .
\ee
Written in the form
\be\label{HJ1.3}
\rho_{\rm tot} (z) = \imath (z- L_{\rm tot})^{-1} \rho_{\rm tot} (t=0) -   \sum_{\tilde \omega}  (z- L_{\rm tot})^{-1}   \delta L \rho_{\rm tot}  (z-\tilde{\omega}) {F}_{\tilde{\omega}} \, .
\ee
\Eq{HJ1.2} is ready for iterating in powers of $\delta L {F}_{\tilde{\omega}}$. But before doing this  we  project to the system and use the separation $t_s\ll t_e$ and arrive, along the lines outlined in Sec.~\ref{insta}, at an equation for the system state $\rho(z)$ incorporating memory and initial correlations with non-Hermitian Liouville $L(z)$ as in \Eq{2.3},
\be\label{HJ1.4}
\rho (z) = \imath (z- L(z))^{-1} \rho_0(z) -   \sum_{\tilde{\omega}} (z- L(z))^{-1}   \delta L \rho (z-\tilde{\omega}) {F}_{\tilde{\omega}} \, .
\ee
Now, we expand  this equation in powers of $\delta L {F}_{\tilde{\omega}}$ and estimate the regime of reliable linear response by requiring that the linear  contribution to $\rho(z)$ is smaller than the contribution without the field.  Since $(z- L(z))^{-1}$ scales {{} with a characteristic system  time, e.g. }$t_{s}$ and the coupling Liouville scales with a characteristic rate as the inverse of a field induced transition time $t_{\rm F_\omega}$, the condition for the linear response theory to  apply reads
\be\label{HJ1.5}
t_{s} \ll t_{\rm  F_\omega} \, .
\ee
In other words, rates of change by the driving field (e.g.~velocities $v$) have to be much smaller than the rate of change by system exploring processes,
\be\label{HJ1.6}
v_{\rm  F_\omega} \ll v_{\rm s} \, .
\ee
Under these conditions we find to linear order
\be\label{HJ1.7}
\rho (z) = \imath (z- L(z))^{-1} \rho_0(z) -   \imath  \sum_{\tilde \omega} (z- L(z))^{-1}   \delta L (z-\tilde{\omega}- L(z-\tilde{\omega}))^{-1} \rho_0 (z-\tilde{\omega}) {F}_{\tilde{\omega}} \, .
\ee
Now we introduce $\xi=z-\tilde{\omega}$ as new variable for later use and write \Eq{HJ1.4} as
\be\label{HJ1.8}
\rho (\xi+\tilde{\omega}) = \imath (\xi + \tilde{\omega}- L(\xi+\tilde{\omega}))^{-1} \rho_0(\xi+\tilde{\omega}) -   \imath  \sum_{\tilde \omega}(\xi + \tilde{\omega}- L(\xi+\tilde{\omega}))^{-1}   \delta L (\xi- L(\xi))^{-1} \rho_0 (\xi) {F}_{\tilde{\omega}} \, .
\ee
We can now use the Laplace transform of the function $f(t)=a\cdot e^{-i\tilde{\omega}t}$ is $f(z)=\imath a /(z-\tilde{\omega})=\imath a /\xi$  to read off the coefficient $a$ of the oscillation with $\tilde{\omega}$ from $\lim_{\xi\to 0} -\imath \xi f(\xi + \tilde{\omega})$,
\be\label{HJ1.9}
\lim_{\xi \to 0} -\imath \xi \rho (\xi+\tilde{\omega}) = 0 -     \sum_{\tilde \omega}(\tilde{\omega}- L(\tilde{\omega}))^{-1}   \delta L \lim_{\xi\to 0}\xi (\xi- L(\xi))^{-1} \rho_0 (\xi) {F}_{\tilde{\omega}} \, .
\ee
As shown in Sec.~\ref{liouv},  $\lim_{\xi\to 0}\xi (\xi- L(\xi))^{-1} \rho_0 (\xi)=\rho_\infty$ and we  arrive at the dynamic susceptibility formula for open systems,
\be\label{HJ1.10}
\chi_{BA}(z) =    -\Tr \LB \LK \LBK G(z) \delta L \RBK \rho_\infty\RK  B \RB , z=\omega +\imath \epsilon\, , \epsilon \to 0^+\, .
\ee
with $G(z)=[z- L(z)]^{-1}$.
By comparing \Eq{HJ1.10} and \Eq{K1.16} we come to a number of conclusions.

\begin{enumerate}
\item The dynamic susceptibility in an open system is formally the same correlation expression as the one given by the Kubo formula. However, the  state $\rho_{\rm eq}$ in the Kubo formula is changed to the stationary state $\rho_\infty$ in the open system, and the resolvent of an isolated system's Liouville $L$ is changed to an open system's relaxator  Liouville $L(z)$. The structure of the susceptibility expression in \cite{WeiYan2011} is also similar to \Eq{HJ1.10}, but the resolvent and state in that expression have to be found by solving a set of hierarchical equations of motion. However, we  can  draw conclusions from the general properties of the relaxator Liouville dynamics.
\item The equilibrium state $\rho_{\rm eq}$ in the Kubo formula  is  the initial state which is unchanged by the isolated  dynamics in the absence of the driving field. In the open system treatment the initial state is not necessarily an equilibrium state, but can be arbitrary, as  the dynamic evolution approaches  a unique (except for non-generic degeneracies) stationary state $\rho_\infty$ in the absence of the driving field.  For  heat bath conditions the stationary state is the canonical equilibrium state $\rho_{\rm eq}$ in accordance with the  Kubo formula. 
\item 
For finite isolated systems the Kubo susceptibility shows singular resonant behavior when the  frequency $\omega$ approaches eigenfrequencies in the limit $\epsilon \to 0^+$. In particular, the dissipative imaginary part vanishes for  almost all frequencies but a number of zero measure. In the case of an open system the limit $\epsilon \to 0^+$ can be performed and the susceptibility stays regular as a function of frequency $\omega$, because  the system contains sufficient damping of isolated resonances  due to environmental contact.
\item The resolvent 
$G(z)$
is  analytic in the upper $z=\omega +i \epsilon$ plane in both cases.
The  Kubo formula is a limiting case of the open system formula when  ${\rm  H} (\omega)\cdot $ reduces to $\LBK H, \cdot\RBK$ and $\Gamma(\omega)$ reduces to constant $\epsilon$ and $\rho_\infty=\rho_{\rm eq}$.
In the open system case the Hermitian  part is odd with respect to $\omega$ and the anti-Hermitian part is even with respect to $\omega$, and 
the operators $\Gamma(\omega)$ and ${\rm  H} (\omega)$ are related to each other by a Kramers-Kronig relation \Eq{dissi4}. 

\item In both cases, the real part ${\chi}'_{BA}(\omega)$ and the imaginary part ${\chi}''_{BA}(\omega)$ of the dynamic susceptibility fulfill a Kramers-Kronig relation  (as a consequence of causality)
\bea
	{\chi}'_{BA}(\omega) = + {\rm  P} \int_{-\infty}^{\infty} \frac{\d \nu }{\pi} \frac{{\chi}''_{BA}(\omega)}{\nu -\omega} \, \label{Kra1}\, , \\
	{\chi}''_{BA}(\omega) = - {\rm  P} \int_{-\infty}^{\infty} \frac{\d \nu }{\pi} \frac{{\chi}'_{BA}(\omega)}{\nu -\omega} \label{Kra2} \, .
\eea
\item The Kubo dynamic susceptibility is an appropriate approximation for  the open system dynamic susceptibility, if
the stationary state $\rho_\infty$ is the equilibrium state $\rho_{\rm eq}$ and  if the average level spacing $\Delta \omega$ of real parts of effective eigenvalues of $L(z)$ as well as the exploring frequency $\omega$ is much larger than the inverse relaxation time $1/\tau_r$ corresponding to $L(z)$. In such regime, now denoted as Kubo regime, the discrete nature of the spectrum of $L(z)$ cannot be resolved by the linear response susceptibility.
In the Kubo regime the resolvent in the system's energy representation (frequency eigenvalues $\omega_{mn}$) can  be replaced by 
\be\label{kvj.1b}
G(\omega)_{mm,nn} = (\omega - \omega_{mn} +\imath /\tau_r)^{-1}\, , 
\ee
with $\tau_r \to \infty$  and dense spectrum of frequency levels $\omega_{mn}$.
 The Kubo regime is adopted to a transport situation,  where a macroscopic system is subject to driving fields (or attached macroscopic contacts with different potentials)  and responding currents occur. The available time to explore the system is $\propto 1/\omega$. In the Kubo regime it  is  much shorter than the time needed to explore the spectral properties of the system, $t_{s}$.
Starting from the Kubo regime, corrections to the Kubo susceptibility  will appear with ${\cal O} \LK \Delta \omega / \omega , \LK \tau_r \cdot \Delta\omega\RK^{-1}\RK $.
\item
A regime, opposite to the Kubo regime, is reached when the frequency $\omega$ can resolve individual level spacings in a finite system,  $\omega \ll \Delta \omega_{\rm s} $, now denoted as regime of isolated levels. Then the open system's dynamic susceptibility probes  individual resonances of the system's propagator  at levels $\omega_k$ and their life times $1/\delta_k$. 
When $\omega$ is close to one of well separated resonances, $\omega=\omega_k(\omega)$, and $\delta_k$ is also smaller than the separation distance  
\be
\chi_{AB}(\omega) \approx f^{AB}_k(\omega) \LK \omega - \omega_k(\omega) + \imath \delta_k(\omega)\RK^{-1}\, \label{kvj.4}
\ee
measures an individual resonance at effective level $\omega_k$ with broadening $\delta_k$ (corresponding to finite life time $1/\delta_k$) and strength factor
\be
f^{AB}_k(\omega)= -\sum_{n,m} \LK{\rho_\infty}_{n} - {\rho_\infty}_{m}\RK \LK AB \RK_{nmk}(\omega)\, .\label{kvj.5}
\ee
Here ${\LK AB \RK}_{nmk}(\omega)$ is a short hand notation,
\be
{\LK AB \RK}_{nmk}(\omega) := \Tr \LB {\sf L}_k^\dagger(\omega) A \RB {{\sf R}_k(\omega)}_{nm} B_{mn} \, .\label{kvj.3}
\ee
In the regime of isolated levels the susceptibilities probe the environmental influence on the frequency spectrum and can help to extract level shifts and finite lifetimes.
\item
$L(\omega)$ defines the time evolution via $G(z)$ in the susceptibility.  $L_H\cdot=[H,\cdot ]$ defines via $\dot{A}=\imath L_H A=\imath [H, A]$ the time derivative of operators within the system. $H$ also defines the canonical equilibrium state. Usually $L(\omega)$ and $L_H$  do not commute and with notable relaxation due to $L(\omega)$  one has to face corrections to the Onsager relations \cite{Onsager}.  When deviating from the Kubo regime corrections to the Onsager relations start with corrections of order ${\cal O} \LK \Delta \omega/\omega , \LK\Delta\omega \cdot \tau_r\RK^{-1} \RK $. 
\end{enumerate}

Dissipation of energy is known to be expressed by the imaginary part ${\chi}''_{BA}(\omega)$. To discriminate the origin of dissipation in the isolated level regime and the Kubo regime, respectively,  we use the matrix representation in the diagonal basis of $\rho_\infty$, 
\be\label{fdt2}
{\chi}''_{BA}(\omega)= \sum_{k;m} \frac{\real\LB \BRA AB \KET_{k;m}(\omega)\RB \cdot \delta_k(\omega)  - \imag \LB \BRA AB \KET_{k;m}(\omega)\RB\cdot \LK \omega -\omega_k(\omega)\RK }{\LK \omega - \omega_k(\omega)\RK^2 + \LK\delta_k(\omega)\RK^2} \, ,
\ee
where  $\BRA AB \KET_{k;m}(\omega) $ is short hand for
\be\label{fdt3}
\BRA AB \KET_{k;m}(\omega):= \rho_{\infty, m} \cdot \Tr\LB L^\dagger_k(\omega) A\RB \cdot \LK \LBK {\sf R}_k(\omega), B\RBK \RK_{mm} \, .
\ee

In the regime of isolated levels in a finite system we put  $\omega$ closely to a resonance, $\omega=\omega_k(\omega)$. 
Then  the dissipation is  proportional to the  lifetime $\tau_k(\omega)=1/\delta_k(\omega)$ of the level, 
\be\label{fdt4aa}
{\chi}''_{BA}(\omega)\approx \LK \sum_{m} \real\LB \BRA AB \KET_{k;m}(\omega)\RB\RK \cdot \tau_k(\omega)  \, .
\ee Thus, relaxing irreversible behavior  characterized by finite lifetimes of eigenfrequencies goes along with dissipative response. 

In the  Kubo  regime the right and left eigenoperators ${\sf R}_k$ and ${\sf L}_k$ are given by dyads of energy eigenstates, $R_{uv}=\mid\left. u \KET \BRA v\mid\right.$ and $L^\dagger_{uv}=\mid\left. v \KET \BRA u\mid\right.$, with $\omega_{uv}=\epsilon_u-\epsilon_v$.
Thus, in the Kubo regime  we have
\bea
{\chi}''_{BA}(\omega)&=&\sum_{uv} \LK \rho_{{\rm eq}, u}-\rho_{{\rm eq} , v}  \RK \LBK \pi \delta\LK \omega -(\epsilon_u -\epsilon_v)\RK \cdot \real \LB A_{uv}B_{vu}\RB \right. \nonumber \\
&-& \left.  
{\rm P}\LK \frac{1}{\omega-(\epsilon_u-\epsilon_v)} \RK \cdot \imag \LB A_{uv}B_{vu}\RB \RBK 
\, .\label{fdt4}
\eea
In particular, for $A=B$, corresponding to a one component field induced dissipation,
\Eq{fdt4} simplifies, by using $\rho_{{\rm eq}, u} = e^{-\epsilon_u/T}/Z$,  to 
\be\label{fdt5}
{\chi}''_{AA}(\omega) = \pi \LK e^{-\omega /T}-1 \RK  \sum_{uv} \rho_{{\rm eq},v} \delta\LK \omega -(\epsilon_u-\epsilon_v) \RK \cdot  |A_{uv}|^2 
 \, .
\ee
Irreversible behavior is thus captured by the Kubo susceptibility, if the spectrum of the Liouville has to be treated  as continuous due to the impossibility to resolve levels with even a small frequency $\omega \gg \Delta \omega_s $. As mentioned before, this is typical for a transport situation where currents enter and leave the system  with scattering boundary conditions or with large reservoirs which cannot - in affordable time - react back to the system's dynamic. In such situation of irreversible behavior in the Kubo regime, time dependent correlation functions typically decay on a characteristic time scale (as discussed in Sec.~\ref{intro})  which we denote as $\tau_{_{BA}}$ for the  time dependent correlation function ${\chi}_{BA}(t)$, the Laplace transform of which defines the susceptibility ${\chi}_{BA}(\omega)$.
A convenient parametrization for the time dependent correlation function $\chi_{BA}(t)$  is given by 
\be\label{fdt6}
{\chi}_{BA}(t) = \imath \LBK BA\RBK_0 e^{- t/\tau_{_{BA}}- \imath t \omega_{_{BA}}}
 \, ,
\ee
where $\LBK BA\RBK_0$ is an initial complex correlation strength and $\omega_{_{BA}}$ captures a possible oscillatory behavior of correlations  with a characteristic frequency $\omega_{_{BA}}$.  
The   time $\tau_{_{BA}}$  is not to be mixed up with the relaxation time $\tau_r$ to reach stationarity in the absence of the driving field. The time $\tau_{_{BA}}$ is due to internal interactions (scattering) yielding relaxation of correlations within affordable times $t\sim 1/\omega$. The  relaxation time $\tau_r$ induced by a coupling to an environment is, in the Kubo regime,   a much larger time scale and  and could be sent to infinity in the end of calculations. 
By  \Eq{fdt6} 
 \be\label{fdt7}
{\chi}_{BA}(\omega) =  -\LBK BA\RBK_0 \LK \omega - \omega_{_{BA}} + \imath /\tau_{_{BA}}\RK^{-1}
 \, .
\ee
Generalizing to frequency dependent parameters $\tau_{_{BA}}(\omega)$ and $\omega_{_{BA}}(\omega)$ allows for a wide range of applications of \Eq{fdt7} (e.g. a discussion of conductivity in weak and strong magnetic fields \cite{GiesHajdu85}, \cite{VIE90}).
 The dissipative part of the dynamic susceptibility reads
 \be\label{fdt8}
{\chi}''_{BA}(\omega) =  \frac{\LB \real \LBK BA\RBK_0 -  \imag \LBK BA\RBK_0  (\omega - \omega_{_{BA}})\tau_{_{BA}} \RB \cdot \tau_{_{BA}} }{ 1  + \LK (\omega - \omega_{_{BA}})\tau_{_{BA}}\RK^2}
 \, .
\ee
In the frequency regime of dominating decay of correlations, $\left.\mid\omega - \omega_{_{BA}}\mid\right.\tau_{_{BA}}\ll 1$, the dissipation is proportional to the decay time of correlations
\be\label{fdt9}
{\chi}''_{BA}(\omega) \approx \real \LBK BA\RBK_0 \cdot \tau_{_{BA}}
 \, .
\ee
As mentioned in  Sec.~\ref{intro} phases of matter  may exist where a dense spectrum of frequencies does not go along with decay of correlation functions and dissipation. In such systems  of  localization with respect to a certain basis (for a review see~\cite{Abanin2019}) the  local spacing between frequencies is huge, states are long-lived and the system cannot dissipate the energy of an external field within affordable time.

To  summarize our findings within  the linear response theory in open systems: there are two distinguishable routes to irreversibility and dissipation: (a) By coupling the system to an environment (e.g. to an external bath or to internal irrelevant variables) discrete frequency levels acquire  finite lifetimes (and frequency shifts) which can be resolved within affordable times by linear response measurements.  The dissipation is proportional to the lifetime of individual levels at resonating conditions (see \Eq{fdt4aa}). The occurrence of lifetimes for discrete levels is caused by a collective action of densely distributed frequency levels of the complementary dynamics.  (b) Inverse lifetimes and spacings of individual levels are to small to be resolved within affordable times  by linear response measurements. Irreversibility and dissipation are related to the collective action of densely distributed frequency levels within the system, leading to decay of correlations within affordable times  of linear response measurements. In the frequency regime of dominating decay of correlations, captured already by the  Kubo formula, the dissipation is proportional to the decay time of correlations (see \Eq{fdt9}). For both routes the reason for irreversibility is the separation of time scales. The affordable time to measure a response of the open system to a perturbation  is very much smaller than the time to explore the frequency content of either the complementary dynamic  or of the system's dynamic.

\section{Weak coupling approximation}\label{weakc}
As indicated at the end of Sec.~\ref{liouv} we now consider an open system weakly coupled to its environment. The weakness condition is related to the fact that we cannot resolve the full spectrum of the environment as well as of the total system. We will neglect interaction energies that are small compared with the affordable resolution scale of $1/t_s$. 
To this end, we decompose  the   eigenfrequencies of $L_\DQ$ as
\bea
E_{m \alpha}-E_{n \beta}&=&\BRA m \alpha \left.\mid \LBK H_\DQ , \mid \right. m \alpha  \KET \BRA n \beta \left. \mid \RBK \right. \mid n \beta  \KET =\nonumber\\
&=&(E_\alpha -E_\beta) + (\epsilon_m -\epsilon_n) + \delta\omega_{m \alpha n\beta } \, ,\label{33.11decomp}
\eea
where $E_\alpha:= \BRA \alpha \left.\mid H_{\rm env} \mid\right. \alpha \KET$, $\epsilon_n:= \BRA n \left.\mid H\mid\right. n \KET$ and 
\be\label{33.decomp2}
\delta\omega_{m\alpha n\beta }= \sum_k   (S_k)_{mm} \LB   (B_k)_{\alpha \alpha} - (B_k)_{\beta\alpha}(\rho_{\rm env})_{\alpha\beta} \RB - (S_k)_{nn}\LB   (B_k)_{\beta \beta} - (B_k)_{\beta\alpha}(\rho_{\rm env})_{\alpha\beta} \RB\, .
\ee
The resolvent of complementary dynamics in the eigenvalue representation is peaked at $\omega=E_{m\alpha}-E_{n\beta}$ and, according to \Eq{33.11decomp}, the energy difference  consists of additive contributions from the environment and the system,
\be\label{decoupling1}
E_{m\alpha}-E_{n\beta }= (E_\alpha-E_\beta) + (\epsilon_m-\epsilon_n)\, ,
\ee
 as long as the contribution from the interaction, $\delta\omega_{\alpha n\beta m}$,  can  be neglected on a resolution scale of order $1/t_{s}$, 
\be\label{decoupling2}
\delta\omega_{\alpha n\beta m} \ll 1/t_{s}\, .
\ee
The approximation, expressed by \Eq{decoupling1} and \Eq{decoupling2}, is denoted as the weak coupling approximation between environment and system.

 In the weak coupling approximation  the  dissipator ${\cal D}(\omega)$  of \Eq{33.11b} can be written as 
\be
{\cal D}(\omega)\rho  =    \sum_{\tilde{\omega}}\sum_{kk'} \LB g_{kk'}(\omega+\tilde{\omega})   \LBK  S_k , (S_{k'} \rho)(\tilde{\omega})\RBK  
-   g^\ast_{kk'}(-\omega-\tilde{\omega})   \LBK  S_k , (S_{k'} \rho)(-\tilde{\omega})\RBK^\dagger  \RB \, , \label{33.14bliou}
\ee
where the environmental correlation function $g_{kk'}(\Omega)$ is defined by 
\be
g_{kk'} (\Omega) := \lim_{\epsilon\to 0^+} \sum_{\alpha, \beta} \frac{\LK \Delta B_k \RK_{\beta \alpha} \LK  \Delta B_{k'} \rho_{\rm env} \RK_{\alpha \beta}}{\Omega - (E_{\alpha } -E_{\beta })+\imath \epsilon} \, ,  \label{33.15aliou} 
\ee
and  the frequency $\tilde{\omega}$ refers to an energy representation of system operators,
\be
X(\tilde{\omega}):=  \sum_{\scriptsize \begin{array}{c}{m, n} \\ {\epsilon_{n} - \epsilon_{ m}=\tilde{\omega}} \end{array}} P_m X P_n=X^\dagger(-\tilde{\omega})\, . \label{33.15aa} 
\ee

By decomposing the denominator in \Eq{33.15aliou} in real and imaginary part,
\be
(\Omega - (E_{\alpha }-E_{\beta })+\imath \epsilon)^{-1} = {\rm P} \frac{1}{ \Omega - (E_{\alpha }-E_{\beta })} -\imath \pi \delta \LK\Omega -(E_{\alpha }-E_{\beta })\RK \label{33.15realima1}\, ,
\ee
we can decompose the correlation function
$g_{kk'}(\Omega)$ accordingly,
\be\label{33.15realima2}
g_{kk'}(\Omega) = s_{kk'} (\Omega) - \imath \gamma_{kk'}(\Omega)  \, .
\ee

The environmental correlation functions $g_{kk'} (\Omega)=s_{kk'}(\Omega)-\imath \gamma_{kk'}(\Omega)$ can be written as
\bea
g_{kk'} (\Omega) &=& \lim_{\epsilon \to 0^+} \BRA  \Delta B_{k} G^{+}_{\rm env}(\Omega) \Delta B_{k'}  \KET_{\rm env} \label{33.16a1}  \, ,\\
s_{kk'} (\Omega) &=& \BRA  \Delta B_{k} {\rm P} (\Omega - L_{\rm env})^{-1} \Delta B_{k'}  \KET_{\rm env} \label{33.16a2} \, ,\\
\gamma_{kk'} (\Omega) &=&  \BRA  \Delta B_{k} \pi \delta(\Omega- L_{\rm env}) \Delta B_{k'}  \KET_{\rm env} \label{33.16a3}=\\
&=& \frac{1}{2}\int_{-\infty}^\infty \d t \, e^{i\Omega t}  \BRA  \Delta B_{k}(t)  \Delta B_{k'}  \KET_{\rm env} \, . \label{33.16a4}
\eea
where  $G^{+}_{\rm env}(\Omega):=(\Omega+\imath\epsilon-L_{\rm env})^{-1}$ is the resolvent with respect to an environmental  dynamics within the weak coupling approximation and $X(t)=e^{\imath L_{\rm env}t} X$ the corresponding Heisenberg operator of $X$.

Note that $\gamma_{kk'}(\omega)$ determines the full $g_{kk'}(\omega)$, since
\be\label{gamma2g}
g_{kk'} (\Omega)=\lim_{\epsilon \to 0^+} \frac{1}{\pi} \int d\Theta \, \frac{\gamma_{kk'}(\Theta)}{\Omega - \Theta +
\imath \epsilon}\, .
\ee
The $s$ correlation function corresponds to $\Delta {\cal H} (\omega)$ and the $\gamma$ correlation function corresponds to $\Gamma(\omega)$, 
\bea
\Delta {\cal H} (\omega) \rho  &=&   \sum_{\tilde{\omega}}\sum_{kk'}  s_{kk'}(\omega+\tilde{\omega}) \LBK  S_k,  (S_{k'} \rho)(\tilde{\omega})  \RBK  -  {s}^\ast_{kk'}(-
\omega -\tilde{\omega}) \LBK  S_k,  (S_{k'} \rho)(-\tilde{\omega})  \RBK^\dagger \label{33.14Liouville1}, \\
\Gamma(\omega) \rho &=&  \sum_{\tilde{\omega}}\sum_{kk'} {\gamma}_{kk'}(\omega+\tilde{\omega}) \LBK  S_k,  (S_{k'} \rho)(\tilde{\omega})  \RBK  
+  {\gamma}^\ast_{kk'}(-\omega-\tilde{\omega}) \LBK  S_k,  (S_{k'} \rho)(-\tilde{\omega})  \RBK^\dagger   .
\label{33.14Liouville2}
\eea

The probability conservation condition $\Tr L(\omega) \rho =0$ is obviously fulfilled in the weak coupling approximation by $L(\omega)$ in \Eq{33.14bliou} due to the commutator structure. In addition we have  $\Tr \Delta {\cal H} (\omega) \rho =0=\Tr \Gamma (\omega) \rho$ due to the commutator structure  in \Eq{33.14Liouville1} and in \Eq {33.14Liouville2}. Furthermore, the conditions (see \Eq{hermi4}, \Eq{hermi5}) $\LK\Delta {\cal H} (\omega) \rho\RK^\dagger=-\Delta {\cal H} (-\omega) \rho$ and $\LK\Gamma (\omega) \rho\RK^\dagger=\Gamma (-\omega) \rho$ are fulfilled.  They keep the time dependent density operator Hermitian. 

For consistency of the weak coupling approximation $\rho_{\rm env}$ is assumed to be  diagonal in the eigenvalue representation, $(\rho_{\rm env})_{\alpha \beta}=(\rho_{\rm env})_\alpha\delta_{\alpha \beta}$. It means that $\rho_{\rm env}$ is not only stationary with respect to $L_{\rm env}$ but also with respect to the complementary dynamics $L_\DQ$  in the weak coupling approximation. As can be inferred from its definition $s_{kk'}(\omega)$ then forms a Hermitian matrix and $\gamma_{kk'}(\omega)$ is a Hermitian and positive matrix,
\bea
 s^\ast_{kk'}(\omega)&=&s_{k'k}(\omega) \, ,\label{skkhermi}\\
 \gamma^\ast_{kk'}(\omega)&=&\gamma_{k'k}(\omega)\, ,\label{gammakkhermi}\\
 \sum_{kk'} \gamma_{kk'}(\omega) v_kv^\ast_{k'}& \geq& 0\, .\label{gammakkposi}
 \eea
 
 With the choice $\rho_{\rm env}=d_{\rm env}^{-1}1_{\rm env}$   $\Delta {\cal H} (\omega)$ and $\Gamma(\omega)$  are still Hermitian super operators in the weak coupling approximation, $\gamma_{kk'}(\Omega)=\gamma_{k'k}(-\Omega)$ and 
$
\Tr  \rho (\Gamma(\omega) \rho)= \Tr  \rho (\Gamma(-\omega) \rho)\geq 0 $ holds true. However this choice is no longer appropriate in the weak coupling approximation as the environment is assumed to be in a self-consistent stationary state $\rho_{\rm env}$ no longer affected by the coupling to the system. Thus, $\rho_{\rm env}$ should be taken as the long time limit stationary state of the environment, e.g. an equilibrium state.
In case the environment persists in thermal equilibrium, described by the  canonical density matrix $\BRA \alpha \right. \mid \rho_{\rm env} \left. \mid \beta \KET = e^{-(E_\beta - F) /T}\delta_{\alpha \beta}$ with  temperature $T$ and canonical free energy  $F$ of the  environment, it is  referred to as the heat bath.  As can be inferred from the definitions \Eq{33.16a2} and \Eq{33.16a3}, the heat bath correlation functions fulfill the Kubo-Martin-Schwinger (KMS) condition of canonical equilibrium (see e.g.
\cite{Haag1967}),
\bea
{s}_{k'k} (-\Omega) &=& -e^{-\Omega/T}s_{kk'} (\Omega) \, , \label{KMS3}\\
\gamma_{k'k} (-\Omega) &=& e^{-\Omega/T}\gamma_{kk'} (\Omega)\, .\label{KMS1}
\eea

In the weak coupling approximation the initial state incorporating correlations, as given by \Eq{inicorham2}, can be written as 
\be\label{inicorham3}
\Delta \rho_0(\omega)=\sum_{\tilde{\omega}} \sum_{k,l} g^0_{kl}(\omega+\tilde{\omega})  \LBK S_k,  \Delta\rho^{(l)}_{{\rm s}0}(\tilde{\omega})\RBK\, ,
\ee
with the environmental correlation function $g^0_{kl}(\Omega)$ 
\be\label{inicorham4}
 g^0_{kl}(\Omega):=\lim_{\epsilon \to 0^+} \sum_{\alpha \beta}\frac{(\Delta B_k)_{\beta\alpha}(\Delta \rho_{{\rm e}0}^{(l)})_{\alpha \beta}}{\Omega+\imath \epsilon -(E_\alpha-E_\beta)}\, .
\ee
The constraint $\Tr \Delta \rho_0(\omega)=0$ is fulfilled in the weak coupling approximation.

As an application of our expression  for the  relaxator Liouville in the weak coupling  approximation \Eq{33.14bliou}  we consider the paradigmatic spin-boson model. The isolated system consists of the smallest possible entity, a two-level system (qubit or spin). The final expression \Eq{300.12} for the relaxator Liouville can  be generalized to arbitrary level number systems by adjusting bare $H$ and the jump-operators $S_k$. It can then be related to other models like matter coupled to quantized radiation fields  in quantum optics (see e.g. Sec.~3.4.1 in \cite{BreuPet}) or to Brownian motion of particles as in the Caldeira Leggett model (see e.g. Sec.~3.6 in \cite{BreuPet} and references therein). The essence of the relaxator Liouville is, in the weak coupling approximation, contained in the correlation functions $\gamma_{kk'}(\Omega)=\BRA  \Delta B_{k} \pi \delta(\Omega- L_{\rm env}) \Delta B_{k'}  \KET_{\rm env} $  of  the environmental coupling operators $B_k$ and the action of the jump-operators $S_k$ on $\rho$, $\LBK S_k, (S_{k'}\rho) (\tilde\omega)\RBK$.

To get explicit expressions  we  considers a  bath of independent bosonic modes of frequency $\omega_\alpha>0$. 
\be\label{300.1}
H_{\rm env}= \sum_\alpha \omega_\alpha a^\dagger_\alpha a_\alpha\, , \, \, \LBK a_\alpha, a_{\alpha'}\RBK =0=\LBK a^\dagger_\alpha, a^\dagger_{\alpha'}\RBK\, ; \LBK a_\alpha, a^\dagger_{\alpha'}\RBK=\delta_{\alpha {\alpha'}}\, .
\ee
The  coupling Hamiltonian is 
\be\label{300.2}
H_{\rm int} =  -S\otimes B\, ,
\ee
where $S$ is called dipole operator of the system and $B$ is a field of bath operators. 

It is tempting to take $B$ linearly in the creation and annihilation operators for simplicity,
\be
B=\sum_\alpha \kappa(\Omega_\alpha) a_\alpha + \kappa^\ast(\Omega_\alpha)a^\dagger_\alpha\, ,
\ee
with coupling strengths  $\kappa(\Omega_\alpha)$.
Such  linear coupling to independent bath modes  is expected to be valid if the relevant wave length $\lambda$ of modes under consideration is much larger than the characteristic size $d$ of the dipole, $\lambda \gg d$. If  characteristic $d$ becomes larger than the relevant $\lambda$, the field  $B$ should  be replaced by  a more general field  involving at least bilinear terms of  creation and annihilation operators. 
Within the linear coupling to independent bath modes one finds 
\be\label{300.5}
\gamma(\Omega) =\pi\nu(|\Omega |)|{\kappa(|\Omega|)}|^2 \cdot   \begin{cases} (1+N(\Omega)) & \mbox{if} \; \Omega > 0 \\ N(|\Omega|) & \mbox{if} \; \Omega < 0 \end{cases}\, .
\ee
Here $N(\Omega)$ is the average equilibrium number of modes {{} and $\nu(|\Omega|)$ the continuous density of frequency levels of bath modes.} However, there is a problematic property of the linear coupling  to independent bath modes: the correlation function $\gamma(\Omega)$ has no characteristic time scale involved besides the time scale inherent in $\rho_{\rm env}$ (e.g. the temperature time scale $T^{-1}$ for environmental equilibrium) {{} and the inverse time scale inherent in the density of levels $\nu(|\Omega|)$ and the inverse time scale set by the  coupling strength $\kappa(|\Omega|)$.} The characteristic time scale of an environmental correlation time $\tau_e$ is missing as an ingredient in a bath of independent modes.  Typically, for independent modes in the limit of large volume  the density of levels  becomes  a power low $\sim \Omega^p$ with $p\geq 1$ such that $\gamma(0) \to 0$. However, for  a realistic  long time correlation function with decay on a scale $\tau_e$ one should find $\gamma(0)\propto \tau_e\not=0$.
Therefore, one shouldn't take the  correlation function of \Eq{300.5} literally, but as a means of parametrization to be adjusted phenomenologically {{} \cite{Weiss}} or to be improved by couplings which allow to introduce characteristic correlation time scales of the environment. {{} Allowing for dependence of the coupling strength on the system's energy levels already makes the linear coupling more reliable for certain aspects of the  dynamics \cite{KarlEtAl}.} For a deeper discussion of  limitations of the linear coupling to independent bath modes we refer to \cite{AlickiQDS} and \cite{Horn2009}. In the following we will not rely on the linear coupling expression \Eq{300.5} but stick to the general properties of $\gamma(\Omega)$ for a bath in canonical equilibrium.

We specify the two-level system  expressed by the  ground state $\left.\mid g\KET$ and the exited state $\left.\mid e\KET$ of system Hamiltonian with excitation energy $\omega_0>0$,
\be\label{300.7}
H=\frac{1}{2}\omega_0 \sigma_3 \, ,
\ee
where $\sigma_3:=P_e-P_g:=\left.\mid e\KET\BRA e\mid\right. - \left.\mid g\KET\BRA g\mid\right.$. For the coupling to the bath modes we take an arbitrary dipole operator, 
\be\label{300.8}
S= S_g P_g + S_e P_e + S_{eg} \sigma_{+} + S_{ge} \sigma_{-}\, ,
\ee
where $\sigma_{+}:=\left.\mid e\KET\BRA g\mid\right.$ excites from the ground state to the excited state and   $\sigma_{-}:=\left.\mid g\KET\BRA e\mid\right.=\sigma_{+}^\dagger$  decreases an excited state to the ground state. Since $S$ is Hermitian we have $S_g,S_e$ real and $S_{ge}^\ast=S_{eg}$.

To evaluate level shift \Eq{33.14Liouville1} and relaxator \Eq{33.14Liouville2} within this setting we have to calculate $\LBK S, (S\rho)(\tilde{\omega})\RBK$. Taking an arbitrary operator $\rho$,
\be \label{300.8b}
\rho=\rho_{g}P_g +\rho_{e} P_e + \rho_{eg}\sigma_{+} +\rho_{ge}\sigma_{-}\, ,
\ee and using the algebra of  $\sigma_{+},\sigma_{-},P_e,P_g$ we find
\bea
\LBK S, (S\rho)(0)\RBK &=&  \LK S_{eg}\sigma_{+}-S_{ge}\sigma_{-}\RK\LBK \LK S_g\rho_g+S_{ge}\rho_{eg} \RK - \LK S_e\rho_e+S_{eg}\rho_{ge}\RK\RBK\, ,\label{300.9}\\
\LBK S, (S\rho)(\omega_0)\RBK &=& \LBK \LK S_g-S_e\RK\sigma_{-} + S_{eg} \sigma_3  \RBK \LK S_g\rho_{ge}+S_{ge}\rho_e\RK\, ,\label{300.10}\\
\LBK S, (S\rho)(-\omega_0)\RBK &=& \LBK \LK S_e-S_g\RK\sigma_{+} - S_{ge} \sigma_3  \RBK \LK S_e\rho_{eg}+S_{eg}\rho_g\RK \, .\label{300.11}
\eea
From Equations~\Eq{300.9}-\Eq{300.11} we can already conclude that, for off-diagonal $S$, the dynamics of populations $\rho_g,\rho_e$ decouples from the dynamics of coherences $\rho_{eg},\rho_{ge}$. For diagonal $S$, the populations are unaffected by the coupling ($H_{\rm int}$ commutes with $H$) and only coherences will change and finally relax to $0$ (pure decoherence).

To write the relaxator Liouville explicitly we consider the bath in canonical equilibrium with temperature $T$, such that the bath correlation function $g(\Omega)=s(\Omega)-\imath\gamma(\Omega)$ fulfills the KMS condition,
\be
g^\ast(-\Omega)= -g(\Omega)e^{-\Omega/T}\, . \label{KMSg}
\ee
The relaxator Liouville in weak coupling \Eq{33.11b1},\Eq{33.14bliou} then reads
\bea
L(\omega)\rho &=& \omega_0 \LK \rho_{eg}\sigma_{+}-\rho_{ge}\sigma_{-} \RK \non \\ &+&
 g(\omega)\LK S_{eg}\sigma_{+}-S_{ge}\sigma_{-}\RK \LB A(S,\rho) -e^{-\omega/T} A^\ast(S,\rho) \RB  \non\\ 
&+&  g(\omega+\omega_0)\LK (S_g-S_e)\sigma_{-}+S_{eg}\sigma_{3}\RK \LB B(S,\rho) -e^{-(\omega+\omega_0)/T} C^\ast(S,\rho) \RB \non\\ 
&-&  g(\omega-\omega_0)\LK (S_g-S_e)\sigma_{+}+S_{ge}\sigma_{3}\RK \LB C(S,\rho) -e^{-(\omega-\omega_0)/T} B^\ast(S,\rho) \RB  \, ,\label{300.12}
\eea
where 
\bea
A(S,\rho)&:=&\LK S_g\rho_g +S_{ge}\rho_{eg}\RK-\LK S_e\rho_e+S_{eg}\rho_{ge}\RK\, ,\label{300.12a}\\
B(S,\rho)&:=& S_g\rho_{ge} +S_{ge}\rho_{e} \, ,\label{300.12b}\\
C(S,\rho)&:=& S_e\rho_{eg} +S_{eg}\rho_{g}\, .\label{300.12c}
\eea
In Equation \Eq{300.12}  $g(\Omega)$ is related to $\gamma(\Omega)$ by \Eq{gamma2g}. 

It is worth stressing that the relaxator Liouville \Eq{300.12} has  the canonical distribution $\rho_{\rm c}=e^{-(H-F)/T}$ with the temperature $T$ inherited from the bath  as its  stationary solution for $\omega\to 0$,
\be 
L(\omega\to 0)\rho_{\rm c}=0\, .\label{300.13} 
\ee
This follows from  ${\rho_{\rm c}}_{eg}={\rho_{\rm c}}_{ge}=0$ and ${\rho_{\rm c}}_{e}={\rho_{\rm c}}_{g}e^{-\omega_0/T}$ for arbitrary $S$. This conclusion is valid due to the weak coupling condition and the KMS condition of the bath, but holds for general couplings $S$ and otherwise arbitrary non-Markovian relaxator Liouville.

In Appendix~\ref{specqubit} we calculate the spectral properties of $L(\omega)$ sufficient for the full time dependence and explicit expressions for the relaxation times of a qubit weakly coupled to a bath from \Eq{300.12}.

\section{Markov approximation: semi-group dynamics}\label{marko}
In the absence of memory and initial correlations the time evolution possesses  the semi-group property, i.e. $\dot{\rho}=-iL\rho$, with $L$ not depending on time nor on frequency and the initial state $\rho_0$ is assumed to be uncorrelated with the environment \cite{AlickiQDS}. The absence of memory   for  relaxator Liouville dynamics means that $L(\omega)$  becomes  independent of $\omega$, at least approximately.  
Weak memory effects correspond to a weakly varying $L(\omega)$. To find a criterion we consider the weak coupling condition \Eq{33.14bliou}  where the frequency dependence is contained in the   environmental  correlation function $\gamma_{kk'}(\omega)$. If its temporal correlation function has a typical range of the environmental correlation time $\tau_e$ then $\gamma_{kk'}(\omega)$ will have a characteristic range of $1/\tau_e$. 
Thus, memory effects are negligible not only in the long time limit but also for  investigation times $1/\omega$ much larger than $\tau_e$.
To have negligible memory effects throughout the whole time evolution of the open system  the  condition
\be\label{cormem1.10}
\tau_e \ll \delta t_{s}\, 
\ee 
has to hold, where $\delta t_{s}$ is the temporal resolution time of the system.
As the Markovian limit of the relaxator Liouville we take 
\be
L:=L(\omega=0)\label{markovlim}
\ee  as a semi-group-generator, because we want to be sure that its stationary limit is as close as possible with that of the original relaxator Liouville.
The condition for the absence of memory  \Eq{cormem1.10}   is quite restrictive and often not fulfilled in a realistic system when times between $\delta t_{s}$ and $\tau_r$ are addressed. One may ask why the semi-group approach that  assumes the absence of memory (and absence of initial correlations) is still quite successful in describing several features of open systems. We think the answer can be given by looking at the solution \Eq{solution} of the relaxator Liouville dynamics in its spectral {{} representation}. The solution \Eq{lastt} shows a superposition of damped oscillations which could as well be described as the solution of a frequency independent Liouville $L$ with its eigenfrequencies adopted to the effective eigenfrequencies $z_k$ \Eq{effeigen} of the full $L(z)$. In particular, when fast oscillatory behavior is averaged out, the Markov approximation, still capturing relaxation, becomes better. 

To make contact with previously known modeling we switch to the Markov limit of our relaxator Liouville within the weak coupling approximation,
expressed by \Eq{33.14Liouville1} and \Eq{33.14Liouville2},  
\be\label{markov1}
L\cdot =L(\omega=0)\cdot=\LBK H_\DP, \cdot\RBK + \Delta {\cal H} (0)\cdot -\imath \Gamma(0)\cdot\, .
\ee
By using the Hermiticity properties of the bath correlation functions and rearrangements of sums we find level shift and relaxator in the Markov limit of the weak coupling approximation, 
\bea
\Delta {\cal H} (0) \rho  &=&   \sum_{\tilde{\omega}}\sum_{kk'} s_{kk'}(\tilde{\omega}) \LK \LBK  S_k,  (S_{k'} \rho)(\tilde{\omega})  \RBK  -  \LBK  (\rho S_{k})(-\tilde{\omega}), S_{k'}  \RBK \RK \, ,\label{markov2} \\
\Gamma(0) \rho &=&  \sum_{\tilde{\omega}}\sum_{kk'} {\gamma}_{kk'}(\tilde{\omega}) \LK \LBK  S_k,  (S_{k'} \rho)(\tilde{\omega})  \RBK  
+  \LBK  (\rho S_{k})(-\tilde{\omega}), S_{k' } \RBK  \RK\, .
\label{markov3}
\eea
To make contact with  the modeling known as the Born-Markov approximation with secular approximation (see e.g. Chap.~3.31 in \cite{BreuPet}) we perform an additional approximation reminiscent of the secular approximation. In a secular approximation, also known as rotating wave approximation, one focuses on those contributions to terms like  $ S_k (S_{k'} \rho)(\tilde{\omega})$ which do not contain strongly fluctuating phase factors and treats them as canceling each other. The secular approximation puts further limitations on the resolution of the time evolution of the density operator as  the typical periods of oscillations are considered to be  small compared to the observational time $t$  before relaxation sets in and such oscillations are averaged out.
To this end we rewrite such terms exactly in the  eigenfrequency representation (see \Eq{33.15aa}) as
\be\label{secular1}
S_k (S_{k'} \rho)(\tilde{\omega})=\sum_{\omega', \omega''} S_k(\omega') S_{k'}(\omega'')\rho(\tilde{\omega}-\omega'')\, .
\ee
Those contributions that fulfill $\tilde{\omega}=\omega''$ and $\omega'=-\omega''$ are free of strongly fluctuating phase factors  and constitute the secular
approximation for \Eq{secular1},
\be\label{secular2}
S_k (S_{k'} \rho)(\tilde{\omega})\approx S_k^\dagger(\tilde{\omega}) S_{k'}(\tilde{\omega})\rho\, ,
\ee
where $\rho(0)$ was replaced by $\rho$. Performing the secular approximation on \Eq{markov2} and \Eq{markov3} we arrive at level shift and relaxator in our weak coupling approximation in the Markov limit with secular approximation,
\bea
\Delta {\cal H}_{\rm sec}  \rho  &=&  \LBK H_{\rm LS}, \rho\RBK\, , H_{\rm LS}:=  \sum_{\tilde{\omega}}  \sum_{kk'} s_{kk'}(\tilde{\omega}) S_k^\dagger(\tilde{\omega})S_{k'}(\tilde{\omega})\, ,\label{markovsecul1} \\
\Gamma_{\rm sec} \rho &=&  \sum_{\tilde{\omega}}\sum_{kk'} {\gamma}_{kk'}(\tilde{\omega}) \LBK \LB S_k^\dagger(\tilde{\omega})S_{k'}(\tilde{\omega}), \rho\RB - 2 S_{k'}(\tilde{\omega})\rho S_k^\dagger(\tilde{\omega})\RBK\, .
\label{markovsecul2}
\eea
These expressions are identical with those of the well known Born-Markov master equation in secular approximation (see e.g. Chap.~3.31 in \cite{BreuPet}). The level shift serves as a simple shift of levels as $H_{\rm LS}$ commutes with $H$ and the relaxator is in Lindblad form and thus generates a so called completely
positive semi-group dynamics. In the diagonal representation of $\rho$ one can show that the relaxator is positive, $\Tr \rho \Gamma_{\rm sec} \rho \geq 0$. 

As an application of the Markovian limit without secular approximation we consider the  Markovian limit of the spin boson model as given in \Eq{300.12}. For the semi-group dynamics of $\rho$ we have
\bea
\dot{\rho}=-\imath L(0)\rho &=& -\imath \omega_0 \LK \rho_{eg}\sigma_{+}-\rho_{ge}\sigma_{-} \RK \non \\ 
   &-& 2 \gamma(0) \LK S_{eg}\sigma_{+}-S_{ge}\sigma_{-}\RK D(S,\rho) \non\\ 
&-& \imath  g(\omega_0)\LK (S_g-S_e)\sigma_{-}+S_{eg}\sigma_{3}\RK \LB B(S,\rho) -e^{-\omega_0/T} C^\ast(S,\rho) \RB \non  \\
&+&  \imath g^\ast(\omega_0) \LK (S_g-S_e)\sigma_{+}+S_{ge}\sigma_{3}\RK \LB B^\ast(S,\rho) -e^{-\omega_0/T} C (S,\rho) \RB  \, ,\label{300.14} 
\eea
where 
\be
D(S,\rho)=S_{ge}\rho_{eg}- S_{eg}\rho_{ge}\, .\label{300.13a}
\ee
Equation \Eq{300.14} leads to linear differential equations for the components of $\rho$,
\bea
\dot{\rho_{e}} &=& 2 \imag \LB g(\omega_0)\LBK a_e(S,T) \LK \rho_e-\rho_{{\rm c},e} \RK + b_e(S,T) \rho_{ge}\RBK  \RB \label{300.15} \\ 
\dot{\rho_{eg}} &=& \LK-\imath \omega_0 +\imath g^\ast(\omega_0)a_{eg}(S,T)\RK\rho_{eg}\non \\
&+& \imath  g^\ast (\omega_0)b_{eg}(S,T) \LK \rho_e-\rho_{{\rm c},e}\RK \non  \\
&-&  2\imath \gamma(0)\LK |S_{eg}|^2 \rho_{ge} - S_{eg}^2\rho^\ast_{eg}\RK\, ,\label{300.16} 
\eea
where 
\bea
a_e(S,T) &:=& |S_{eg}|^2 \LK1+e^{-\omega_0/T}\RK\, , \;\; b_e(S,T) := S_{eg} \LK S_g -S_e e^{-\omega_0/T}\RK   \, ,\label{300.17}\\
a_{eg}(S,T) &:= &\LK S_g-S_e\RK \LK S_g -S_e e^{-\omega_0/T}\RK \, , \; b_{eg} (S,T) := \LK S_g-S_e\RK S_{eg}\LK1+ e^{-\omega_0/T}\RK  ,\label{300.18}
\eea
and 
\be\label{300.19}
\rho_{{\rm c},e}=\frac{e^{-\omega_0/T}}{1+e^{-\omega_0/T}}
\ee is the equilibrium canonical population in the excited state.

From Equations \Eq{300.15}-\Eq{300.18} we conclude for the qubit weakly coupled  to a heat bath in the Markov approximation: as soon as there is some off-diagonal coupling,  $S_{eg}\not=0$, which means  excitation and emission processes, the state will relax to the canonical equilibrium state with relaxation time 
\be\label{300.20}
\tau_r = \LBK 2\gamma(\omega_0) |S_{eg}|^2 \LK1+e^{-\omega_0/T}\RK\RBK^{-1} \,.
\ee
The dying out of the coherence  $\rho_{eg}$  depends on  the diagonal elements of $S$ and in the generic case of a  non-vanishing  $\gamma(0)$ also on the off-diagonal elements. In case off a purely diagonal $S$, the populations are unaffected by symmetry and the coupling leads to a scenario of pure decoherence with decoherence time 
\be\label{300.21}
\tau_{\rm dec} = \LBK \gamma(\omega_0)\LK S_g-S_e\RK  \LK S_g-S_e e^{-\omega_0/T}\RK\RBK^{-1} \,.
\ee
 In the non-generic case of purely off-diagonal $S$ and vanishing $\gamma(0)$ we would have a degenerate asymptotic state: relaxation of populations to equilibrium with relaxation time $\tau_r$ and oscillations of coherences with frequency $\omega_0$. In a secular approximation calculation such oscillations cannot be  captured.  
 
\section{Summary {{} and discussion}}\label{summa}
We developed the relaxator Liouville dynamics (Eqs.~(\ref{Int4}),(\ref{2.3}) - (\ref{dissi4})) for relevant degrees of freedom in a macroscopic system from a microscopic von Neumann equation for an isolated total system of all degrees of freedom. The irrelevant degrees of freedom act as an environment on the relevant degrees of freedom. The relaxator $\Gamma(\omega)$ indicates the onset of irreversibility. It can be  expressed by a Hamiltonian modeling of system environment coupling (Eq.~(\ref{GammaInH})). For a weak coupling  the relaxator Liouville is determined by environmental correlation functions and hopping operators in the system (Eq.~\Eq{33.14bliou}).
The relaxator Liouville dynamics has generically a unique stationary state (Eq.~\Eq{3.4}) which is reached after a characteristic relaxation time that can be calculated from the relaxator Liouville's effective eigenvalues  (Eq.~\Eq{lastt2}). For equilibrium conditions the system reaches canonical thermal equilibrium with a temperature adopted to the environment.

We derived a closed kinetic equation for the one-particle density operator (Eq.~(\ref{kin8})) as a special form of relaxator Liouville dynamics where one-particle degrees of freedom are the relevant ones. The origin of irreversible behavior can be identified via the relaxator (Eq.~(\ref{kin21})). 

For situations of current carrying stationary states in the presence of perturbing external fields we derived linear response expressions for dynamic susceptibilities (Eq.~\Eq{HJ1.10}). The derived susceptibilities coincide with Kubo's susceptibilities in a regime where the energy eigenstates  of the unperturbed system have undetectable large lifetimes and the probing frequency is much larger than the typical level spacing.  Despite the practically infinite lifetimes of energy eigenstates, the response can show an collective irreversible character as the spectrum of the system becomes dense relative to the frequency.  The corresponding  time scale of irreversibility, a decay time of correlations within the system, can be extracted from the susceptibility (Eq.~(\ref{fdt9})). {{} In the limit of separated system levels with finite life times the linear response susceptibility allows to extract lifetimes and level shifts (Eq.~(\ref{kvj.4})).}

We identified Markov semi-group dynamics as limiting modelings of our more general relaxator Liouville dynamics (\Eq{markov2},\Eq{markov3}).

As the source of irreversibility we have identified time scale separations. For a system in contact with a surrounding bath the time to {{} resolve}  the spectrum of the bath is much larger than the time to {{} resolve} the spectrum of the isolated system. For a macroscopic system in general the dynamics of the so-called relevant variables becomes irreversible when the time to {{} resolve} the spectrum of isolated relevant variables is much smaller than the time to {{} resolve} the spectrum of the total system. In transport situations with driving fields, irreversibility and dissipation of energy emerges when they are macroscopic or subject to macroscopic contacts such that the level spacing of the system cannot be resolved with the frequency of the applied field. In other words: the time to {{} resolve} the system's spectrum is much larger as any  time we can afford to measure the response of the system to applied fields.

{{}
Our general framework for modeling irreversible systems  offers a  generalization  of the  von Neumann dynamics for isolated relevant degrees of freedom to  a relaxator Liouville dynamics of relevant degrees of freedom in open systems. It captures non-Markovian dynamics by the frequency dependence of $L(\omega)$. The total Hamiltonian is however assumed to be time independent (except for the investigation of linear response in Chap.~\ref{linea}) and thus the total energy is conserved. Degrees of freedom are permanently treated as relevant or as irrelevant. Under these restrictions the full time dependence of environmental correlation functions can enter the construction of $L(\omega)$ and of $\rho_0(\omega)$ as given by Eqs. \Eq{2.3}-\Eq{dissi4} and \Eq{33.8bliou}-\Eq{inicorham2}. The resolution of $L(\omega)$ is assumed to be reliable down to a frequency resolution of the typical level spacing of isolated relevant variables, $1/t_s$. Since this allows for non-Markovian dynamics including initial correlations with the environment a detailed time evolution of relevant degrees of freedom is possible. However this framework is aimed at and adopted to a situation where the number of irrelevant (environmental) degrees of freedom is huge compared to the number of relevant degrees of freedom. Due to the spectral condition ($t_s << t_e $ (the level density of the total system cannot be resolved on the time scale $t_s$) the individual interaction processes between system and environment and within the environment are fast and the time evolution of such individual processes cannot be resolved in detail. If both $t_s$ and $t_e$ are of the same order the whole approach of a relaxator Liouville does no longer make sense. Both sets of variables (relevant and irrelevant) then have to be treated on the same level of resolution as can be done by the methods discussed in \cite{HoffEtAl}.  In \cite{HoffEtAl} a cavity of two levels interacting with 400 photon modes is considered  to describe precisely the emission, reflection, absorption and re-emission with interference of a one-photon wave packet. Such detailed interaction dynamics  is beyond the scope of a Liouville relaxator modeling and  other methods as investigated in \cite{HoffEtAl} are surely superior. For $t_e \sim t_s$ is is hard to reach relaxation to steady state situations. The domain  of applicability  of a relaxator Liouville modeling is  $t_s << t_e$. It is adjusted for studying the non-Markovian approach to long time dynamics and steady state situations. With available spectral representation of $L(\omega)$  the full dynamics of relevant  variables  is given by \Eq{lastt}.  Whenever it comes to the irreversible time evolution of relevant degrees of freedom due to many individual interaction processes captured statistically by means of a relaxator $\Gamma(\omega)$ the relaxator Liouville modeling device is general and  flexible. 
So far, the construction of $L(\omega)$ is straightforward in the weak coupling approximation where  $\Gamma(\omega)$ separates into bath correlation functions and hopping operators in the system. To go beyond the weak coupling approximation by systematic methods that keep the total energy conserved is left for further investigations as well as applications to specific systems.}
\medskip

\noi\textbf{Acknowledgements}  
 The authors acknowledge valuable comments during the process of writing from Victor Albert (2018),  Heinz-Peter Breuer (2018) , John Chalker (2024),  Rochus Klesse (2017),  Dimitry Polyakov (2025),  J\'anos Polonyi (2017, 2025) and Peter Wölfle (2025). {{} The authors acknowledge support by the University of Cologne for providing open acces for publication in AdP.}

\noi{Martin Jan{\ss}en  would like to express his deep gratitude for having shared many years with the late co-author János Hajdu. He was an  instructive  scholar and a  wonderful person. As a scholar, he was distinguished by his comprehensive erudition and pedagogical skills, and as a person,  by his great attentiveness and helpfulness.}

\medskip
\noi\textbf{Conflict of Interest}  The authors declare no conflict of interest.

\medskip
\noi\textbf{Data Availability Statement} Data sharing is not applicable to this article as no new data were created
or analyzed in this study.

\begin{appendix}
\section{Approaching  canonical equilibrium in weak coupling to a bath}\label{robustenvmod}
We consider the relaxator  within the weak coupling approximation  \Eq{markov3} for $\omega=0$ and calculate its elements $\Gamma_{rr,nn}$ according to 
\Eq{comp1}
\bea
\Gamma_{rr,mm}(\omega=0)&=&
\LK   \BRA r| \Gamma(0)P_n|r\KET\RK =   \non\\ 
&=& \sum_{\tilde{\omega}}  \sum_{k,k'} \gamma_{kk'}(\tilde\omega)\LB \BRA r \left.\mid\LBK S_k, \LK S_{k'}P_n\RK (\tilde{\omega})\RBK + 
\LBK \LK P_nS_{k}\RK (-\tilde{\omega}), S_{k'}\RBK\mid\right. r\KET\RB\, .\label{333.1}
\eea
Since $P_n$ commutes with other $P_q$, the eigenfrequency representations in \Eq{333.1} read $\LK S_{k'}P_n\RK(\tilde{\omega})= S_{k'}(\tilde{\omega}) P_n$ and $\LK P_n S_{k}\RK(-\tilde{\omega})= P_n S_{k}(-\tilde{\omega})$. For $n\not= r$ we find the restriction  $\tilde{\omega}=\epsilon_n-\epsilon_r$ in 
\be
\Gamma_{rr,mm}(\omega=0) =  - \sum_{\tilde{\omega}}  \sum_{k,k'} \gamma_{kk'}(\tilde\omega)\LB \BRA r|S_k|n\KET \BRA n| S_{k'}|r\KET + \BRA r|S_{k'}|n\KET \BRA n| S_{k}|r\KET \RB .\label{333.2}
\ee
The resulting transition rate $W_{rn}=-\Gamma_{rr,nn}(\omega=0) \geq0$ ($n\not=r$) due to the positive matrix $\gamma_{kk'}$.
By the KMS condition \Eq{KMS1} on the bath correlation function $\gamma_{kk'}(\tilde{\omega})$ with $\tilde{\omega}=\epsilon_n-\epsilon_r$  for the bath in canonical equilibrium at temperature $T$ we conclude 
\be
W_{nr}=e^{-(\epsilon_n- \epsilon_r)/T}W_{rn} \, ,\label{333.3}
\ee
which is the condition \Eq{caneq} for reaching canonical equilibrium in the system with the bath temperature $T$.
\section{qubit  weakly coupled to a bath: spectral properties and relaxation times} \label{specqubit}
As a prototypical example for non-Markovian dynamics we consider the relaxator Liouville of a qubit weakly coupled to a bath  as given by \Eq{300.12}. The time evolution  is, according to \Eq{lastt2},  characterized by   the effective eigenvalues $z_k$,  the stationary state and coefficients depending on the initial state. We look at two special cases: (I) The coupling operator $S$ is purely diagonal and (II) the coupling operator $S$ is purely off-diagonal. 

In  case (I) $S=S_eP_e+S_g+P_g$ commutes with the system Hamiltonian $H=\frac{1}{2}\omega_0(P_e-P_g)$. This  leads to pure decoherence as the populations $\rho_e=1-\rho_g$ are not affected by the dynamics. Every diagonal density operator $\rho=\rho_e(P_e -P_g) + P_g$ is a right eigenoperator  of $L(\omega)$ with vanishing eigenvalue,  $\lambda=0$. The eigenvalue $0$ is twofold degenerate. As two $\lambda=0$ eigenoperators  of a bi-orthonormal set of $4$ eigenoperators, we can choose:
\be\label{specqu1} 
{\sf R}_0 = \rho_e(P_e -P_g) + P_g \, , {\sf R}_1= \sigma_3\, .
\ee
The corresponding left eigenoperators follow from the bi-orthonormal properties as
\be\label{specqu2}
 {\sf L}_0 = 1 \, , {\sf L}_1= -\rho_e(P_g+P_e)+P_e\, .
\ee
We can choose $\rho_e$ in \Eq{specqu1} as the initial population in the excited state. Equation \Eq{lastt2} confirms  that the populations stay constant during the time evolution. 
From Equation \Eq{300.12} for  case (I) we conclude that excitation and absorption form the two remaining right eigenoperators,
\be\label{specqu3}
{\sf R}_2= \sigma_{+}\, , {\sf R}_3=\sigma_{-}\, . 
\ee
By comparison with the eigenvalue condition we find the corresponding eigenvalues,
\bea
\lambda_2(\omega) &=& \omega_0 +g(\omega-\omega_0)\LK S_e-S_g\RK\LK S_e-S_g e^{-(\omega-\omega_0)/T}\RK=:\lambda_{+}(\omega) \, , \non \\ \lambda_3(\omega)&=& -\omega_0 +g(\omega+\omega_0)\LK S_g-S_e\RK\LK S_g-S_e e^{-(\omega+\omega_0)/T}\RK=:\lambda_{-}(\omega) \label{specqu4}\, .
\eea
The left eigenoperators follow from the bi-orthonormal properties as
\be\label{specqu5}
{\sf L}_2= \sigma_{+}={\sf R}_2 \, , {\sf L}_3=\sigma_{-}={\sf R}_3\, . 
\ee
The relaxation of the coherences is described by a damped oscillation characterized by two effective complex eigenvalues 
\be\label{specqu6}
z_k=\omega_k-\imath\delta_k =\lambda_k(\omega_k)\, , k=+,-\, ,
\ee
where Hermiticity of $\rho(t)$ requires 
\be\label{specqu7}
z_{+}=-z^\ast_{-} \, .
\ee
The real part, $\omega_{+}=-\omega_{-}$,  is the solution of the real part of \Eq{specqu6} involving $s(\omega_{+} - \omega_0)$, describing the oscillatory behavior. The relaxation is described by $\delta_{+}=\delta_{-}$ and involves $\gamma(\omega_{+} - \omega_0)$. $\delta_{+}$ sets the decoherence time scale,
\be\label{specqu8}
\tau_{\rm dec}=  \LBK \gamma(\omega_{+}- \omega_0)\LK S_e-S_g\RK\LK S_e-S_g e^{-(\omega_{+} -\omega_0)/T}\RK \RBK^{-1}\, .
\ee 
Since $s(0)=0$ holds for symmetry reasons, the solution $\omega_{+}=\omega_0$ solves the real part of \Eq{specqu6} and we find 
\be\label{specqu9}
\tau_{\rm dec}=\LBK \gamma(0)\LK S_e-S_g\RK^2\RBK^{-1}\, .
\ee 
A vanishing of $\gamma(0)$ would prohibit  decoherence. However, as discussed before, a  vanishing $\gamma(0)$ is  unrealistic  for a heat bath.  Equation \Eq{specqu8} in comparison to  \Eq{300.21}  shows a  characteristic difference between the non-Markovian relaxator Liouville dynamics and the  Markov limit to that dynamics: while in the Markov case the limit $\omega\to 0$ comes first and  then the eigenvalue of $L(0)$  counts, in the non-Markovian approach the effective eigenvalue of the frequency dependent relaxator Liouville counts. A difference between the  decoherence times  of \Eq{specqu8} and\Eq{300.21} will not be detectable if $\omega_0$ is much smaller than the inverse of the environmental correlation time $\tau_e$ (see the discussion in Sec.~\ref{marko}).

Switching to  case (II), $S=S_{ge}\sigma_{-}+S_{eg}\sigma_{+}$ we write the relaxator Liouville of \Eq{300.12} as
\bea
L(\omega)\rho &=& \omega_0 \LK \rho_{eg}\sigma_{+}-\rho_{ge}\sigma_{-} \RK \non \\ &+&
 g(\omega)\LK  S_{ge}\rho_{eg}-S_{eg}\rho_{ge}\RK  \LK 1+e^{-\omega/T}\RK \LK S_{eg}\sigma_{+}-S_{ge}\sigma_{-}\RK   \non\\ 
&+&  |S_{eg}|^2 \LB g(\omega+\omega_0)\LK  \rho_e-\rho_g e^{-(\omega+\omega_0)/T} \RK - g(\omega-\omega_0)  \LK  \rho_g -\rho_e e^{-(\omega-\omega_0)/T} \RK\RB \sigma_{3} \, .\non \\ \label{specqu10}
\eea
The $\lambda_0(\omega)=0$ right eigenoperator with unit trace can be found by looking for $\rho_e$ for which the factor in front of $\sigma_3$ in \Eq{specqu10} vanishes.  Obviously $\sigma_3$ is a right eigenvector with non vanishing eigenvalue $\lambda_1(\omega)$. Thus, we have
\be\label{specqu11} 
{\sf R}_0(\omega) = \rho_e(\omega) (P_e -P_g) + P_g \, , \lambda_0(\omega)=0\, , {\sf R}_1(\omega)= \sigma_3\, ,
\ee
with
\be\label{specqu12}
\rho_e(\omega)=\frac{g(\omega+\omega_0)e^{-(\omega+\omega_0)/T}+g(\omega-\omega_0)}{g(\omega+\omega_0)\LK 1+e^{-(\omega+\omega_0)/T}\RK +g(\omega-\omega_0)\LK  1+e^{-(\omega-\omega_0)/T}\RK}\, .
\ee
The effective eigenvalue $z_0=0$ requires $\omega_{k=0}=0$ and at this value we have the stationary canonical equilibrium density operator with
\be\label{specqu13}
\rho_{e,\infty}=\rho_e(0)=\frac{e^{-\omega_0/T}}{1+e^{-\omega_0/T}}=\rho_{{\rm can}, e}\, .
\ee
The eigenvalue $\lambda_1(\omega)$ can be seen from the prefactor of $\sigma_3$ in \Eq{specqu10} and reads
\be\label{specqu14}
\lambda_1(\omega)=|S_{eg}|^2\LB g(\omega +\omega_0)\LK 1+e^{-(\omega+\omega_0)/T}\RK+ g(\omega -\omega_0)\LK 1+e^{-(\omega-\omega_0)/T}\RK\RB\, .
\ee
The real part of the effective eigenvalue equation yields a vanishing frequency $\omega_1=0$ but a finite $\delta_1$ which defines the relaxation time for the populations,
\be\label{specqu15}
\tau_{r, {\rm diag}}=\LBK 2|S_{eg}|^2\gamma(\omega_0)\LK 1+e^{-\omega_0/T}\RK \RBK^{-1}\, .
\ee
This unsurprisingly coincides with the result in the Markov limit \Eq{300.20}.

The off-diagonal right eigenoperators and corresponding eigenvalues can be found by a standard matrix eigenvalue routine for   $L(\omega) \rho$ in the $\sigma_{\pm}$ basis. We find the eigenvalues as
\be\label{specqu16}
\lambda_{2,3}(\omega)=f(\omega,T,S) \pm \LK \omega_0 +\delta z_0(\omega,T,S) \RK\, ,
\ee
where 
\bea
f(\omega,T,S):=g(\omega)\LK 1+e^{-\omega/T}\RK |S_{ge}|^2\, , \label{specqu17}\\
\omega_0 + \delta z_0(\omega,T,S):=\omega_0\LK 1+ \frac{|f|^4}{\omega_0^4} + 2\frac{|f|^2\cos(2\varphi)}{\omega_0^2}\RK^{1/4}e^{-\imath\tilde{\varphi}/2}\, ,\label{specqu18}\\
\tan(\varphi):=\frac{s}{\gamma} \, , \tan(\tilde{\varphi}):=\frac{|f|^2\sin(2\varphi)}{\omega_0^2+|f|^2\cos(2\varphi)} \, . \label{specqu19}
\eea
The right eigenoperators read 
\bea 
{\sf R}_2(\omega) = \frac{1}{\LK 1+|\frac{\delta z_0}{\omega_0+f}|^2\RK^{1/2}}\LK \sigma_{+} + \frac{\delta z_0}{\omega_0+f} \sigma_{-}\RK\, ,\label{specqu20a} \\
{\sf R}_3(\omega) = \frac{1}{\LK 1+|\frac{\delta z_0}{\omega_0+f}|^2\RK^{1/2}}\LK \sigma_{-} - \frac{\delta z_0}{\omega_0+f} \sigma_{+}\RK\, .\label{specqu20b}
\eea
The corresponding left eigenoperator $L_2=c_2\sigma_{+}+d_3\sigma_{-}$ for two given right eigenoperators $R_{2,3}=a_{2,3}\sigma_{+}+b_{2,3} \sigma_{-}$ can be constructed by the following scheme due to the bi-orthonormal properties
\be
c^\ast_2=\frac{b_3}{a_2b_3-a_3b_2}\, , \, d^\ast_2=\frac{-a_3}{a_2b_3-a_3b_2}\, .
\ee\label{specqu21}
The effective eigenvalues $z_2=-z_3^\ast$ are found by solving the real part equation 
\be
\omega_2= \omega_0 + \real \LK f(\omega_2) + \delta z_0(\omega_2)\RK\label{specqu22a}
\ee and calculating the imaginary part that defines the relaxation time for coherences,
\be
\label{specqu22}
\tau_{\rm dec}=\LBK \gamma(\omega_2)(1+e^{-\omega_2/T})|S_{eg}|^2 + \imag \delta z_0(\omega_2) \RBK^{-1}\, .
\ee
Note, that here the decoherence time is different from the Markov limit value as $\omega_2$ is shifted from $\omega_0$. For weak coupling the shift amounts to a perturbation of ${\cal O} (|f(\omega_0)|/\omega_0)$ and for short environmental correlation times $\tau_e\ll \omega_0^{-1}$ the difference in decoherence times will  be negligible.
\end{appendix}

\medskip


\begin{thebibliography}{99}
\bibitem{Zwiebach2022} B. Zwiebach, {\em Mastering Quantum Mechanics: Essentials, Theory, and Applications}, MIT Press, Cambridge, MA (2022).
\bibitem{LL5} L. D. Landau, E.  M  Lifshitz, {\em Statistical Physics. Vol. 5 (3rd ed.)}, Butterworth-Heinemann,  (1980).
\bibitem{Boltzmann1872} L Boltzmann, {\em Weitere Studien \"uber das W\"armegleichgewicht unter Gasmolek\"ulen}, in: \journal{Sitzungsberichte der Kaiserlichen Akademie der Wissenschaften} {66}{275}{1872}, reprinted in L. Boltzmann, {\em Wissenschaftliche Abhandlungen, Bd. 1., p. 316}, J. A. Barth, Leipzig (1909).
\bibitem{Smoluchowski1916} M.v. Smoluchowski, {\em Drei Vortr\"age \"uber Diffusion, Brownsche Bewegung und Koagulation von Kolloidteilchen}, \journal{Physik. Zeit.}{17}{557}{1916}.
\bibitem{Pauli} W. Pauli, {\em Festschrift zum 60. Geburtstage A. Sommerfelds, p.30} Hirzel, Leipzig (1928).
\bibitem{WeisWig} V.~Weisskopf and E.~Wigner, {\em Berechnung der nat\"urlichen Linienbreite auf Grund der Diracschen Lichttheorie}, \journal{Z. Phys.}{63}{54}{1930}; {\em \"Uber die nat\"urliche Linienbreite in der Strahlung des harmonischen Oszillators}, \journal{Z. Phys.}{65}{18}{1930}.
\bibitem{vanHove} L.v. Hove, {\em Quantum-Mechanical Perturbations Giving Rise to a Statistical Transport Equation}, \journal{Physica}{21}{517}{1955}, \journal{Physica}{23}{441}{1957}, \journal{Physica}{25}{268}{1959}.
\bibitem{Kubo1957} R. Kubo, {\em Statistical Mechanical Theory of Irreversible Processes. I.} \journal{J. Phys. Soc. Japan}{12}{570}{1957}.
\bibitem{Prigogine1963} I. Prigogine, {\em Non-Equilibrium Statistical Mechanics}, Interscience Publishers, New York, (1963).
\bibitem{Zwanzig2001} R. Zwanzig, {\em Non Equilibrium Statistical Mechanics}, Oxford University Press, Oxford, (2001).
\bibitem{Uffink2007} J. Uffink, {\em Compendium of the foundations of classical statistical physics}, in:  Jeremy Butterfield \& John Earman (eds.), {\em Handbook of the Philosophy of Physics, p.~923},  Elsevier, (2007).
\bibitem{Batalhao2018}T.B. Batalhão, S. Gherardini, J.P. Santos, G.T. Landi, and  M. Paternostro, {\em Characterizing Irreversibility in Open Quantum Systems}, in: 
F. Binder, L.A. Correa, C. Gogolin, J. Anders, G. Adesso (eds.), {\em Thermodynamics in the Quantum Regime}, Springer, Cham, (2018), arXiv:1806.08441 [quant-ph].
\bibitem{Roberts2022} B.W. Roberts, {\em Reversing the Arrow of Time}, Cambridge University Press, Cambridge, (2022), DOI 10.1017/9781009122139.
\bibitem{Handbook2011} C. Callender (ed.), {\em The Oxford Handbook of Philosophy of Time}, Oxford University Press Inc., New York (2011).
\bibitem{Ven2015} L.C. Venuti, {\em The recurrence time in quantum mechanics},   arXiv:1509.04352v2 [quant-ph] (2015).
\bibitem{Zia2009} R.K.P. Zia, E.F. Redish, and S.R. McKay, {\em Making Sense of the Legendre Transform}, arXiv:0806.1147 [physics.ed-ph] (2009).
\bibitem{Nakajima1959}S. Nakajima, {\em On Quantum Theory of Transport Phenomena}, \journal{Prog. Theor. Phys.}{20}{948}{1958}.
\bibitem{Zwanzig1960} R. Zwanzig, {\em Ensemble Method in the Theory of Irreversibility}, \journal{J. Chem. Phys.}{33}{1338}{1960}.
\bibitem{Fano} U.~Fano, {\em Liouville Representation of Quantum Mechanics with Application to Relaxation Processes} in: E.~R.~Caianiello, 
{\em Lectures on the Many Body Problem Vol.~2}, p.~217, ~Academic~Press,~New~York,~(1964).
\bibitem{Abanin2019} D.A. Abanin, E. Altman, I. Bloch, and M. Serbyn, {\em Colloquium: Many-body localization, thermalization, and entanglement}, \journal{Rev. Mod. Phys.}{91}{021001}{2019}.
\bibitem{BreuEtal} H.P. Breuer, E.-M. Laine, J. Piilo, and B. Vacchini {\em Non-Markovian dynamics in open quantum system}, \journal{Rev. Mod. Phys.}{88}{021002}{2016}, arXiv:1505.01385v1[quant-ph].
\bibitem{Maldonado-MundoEtAl2012} D. Maldonado-Mundo, P. \"Ohberg, B. W. Lovett, and E. Andersson, {\em Investigating the generality of time-local master equations},
\journal{Phys. Rev. A}{86}{042107}{2012}, DOI https://doi.org/10.1103/PhysRevA.86.042107.
\bibitem{ZhangEtAl2012} W.-M. Zhang, P.-Y. Lo, H.-N. Xiong, M. W.-Y. Tu, and F. Nori, {\em General Non-Markovian Dynamics of Open Quantum Systems}, \journal{PRL}{109}{170402}{2012}.
\bibitem{Bonitz2016} M. Bonitz, {\em Quantum Kinetic Theory (2nd ed.)}, Teubner-Texte zur Physik, Springer, Cham, (2016).
\bibitem{VLeuwen2006} R. van Leeuwen and N.E. Dahlen {\em An Introduction to Nonequilibrium Green Functions}, lecture of  3rd International Workshop and School on Time-Dependent Density-Functional Theory: Prospects and Applications, https://www.tddft.org/TDDFT2008/lectures/RL\_LN.pdf (2006) (accessed 2023/07/24).
\bibitem{Datta} K.Y. Camsari, S. Chowdhury and S. Datta, {\em The Non-Equilibrium Green Function (NEGF) Method}, in: Springer Handbook of Semiconductor Devices, p.~1583, Springer, (2023),  	  arXiv:2008.01275v2 [cond-mat.mes-hall].
\bibitem{Lindblad} G. Lindblad, {\em On the generators of quantum dynamical semi-groups}, \journal{Commun. Math. Phys.}{48}{119}{1976}.
\bibitem{Gorini} V. Gorini, A. Kossakowski and E.C.G. Sudarshan, {\em Completely Positive Dynamical Semigroups of N Level Systems}, \journal{Journ.
of Math. Phys.}{17}{821}{1976}.
\bibitem{BreuPet} H.P. Breuer and F. Petruccione, {\em The Theory of Open Quantum Systems}, Oxford University Press, Oxford, (2002).
\bibitem{AlickiQDS}  R. Alicki, {\em Invitation to quantum dynamical semi-groups}, arXiv:0205188[quant-ph] (2002).
\bibitem{BreuPet2003} H. P. Breuer, and F. Petruccione, {\em Concepts and methods in the theory of open
quantum systems} in:  F.Benatti, and R.Floreanini (Eds.), {\em Irreversible Quantum Dynamics}, p.~pp 65, Lecture Notes in Physics, Vol. 622, Springer, Berlin, Heidelberg, (2003), 	
arXiv:quant-ph/0302047[quant-ph].
\bibitem{RivasHuelga12} A.~Rivas, and F.S. Huelga, {\em Open Quantum Systems. An Introduction}, Springer Berlin, Heidelberg,  arXiv:1104.5242v2 [quant-ph] (2012).
\bibitem{Lidar2020} D. Lidar, {\em Lecture Notes on the Theory of Open Quantum Systems}, arXiv:1902.00967v2 (2020).
\bibitem{Jan2018} M. Jan{\ss}en, {\em Equation of Motion for Open Quantum Systems incorporating Memory and Initial Correlations},  arXiv:1810.06458v1[quant-ph] (2018).
\bibitem{Pol2011} J. Polonyi,  {\em Dynamical breakdown of time reversal invariance and causality}, \journal{Phys. Rev.  D}{84}{105021}{2011}, arXiv:1109.2228 [hep-th].
\bibitem{JanBook2016} M. Jan{\ss}en, {\em Generated Dynamics of Markov and Quantum Processes}, Springer-Verlag, Berlin Heidelberg, (2016).
\bibitem{YoshiSa2025} T. Yoshimura, and L. Sa, {\em Theory of Irreversibility in Quantum Many-Body Systems}, arXiv:2501.06183v3 [cond-mat.stat-mech] (2025).
\bibitem{vKamp} N.G. van Kampen, {\em Stochastic Processes in Physics and Chemistry, 2nd edn.}, North-Holland,Amsterdam, (1992).
\bibitem{VKampInNoise} N.G. van Kampen,  {\em Internal Noise}, Chapter XVII 7 in \cite{vKamp}.
\bibitem{Jan2017} M. Jan{\ss}en, {\em On Generated Dynamics for Open Quantum Systems: Spectral Analysis of Effective Liouville},  arXiv:1707.09660[quant-ph] (2017).
{{} \bibitem{StefAlm}  G. Stefanucci, and C.O. Almbladh, {\em Time-Dependent Partition-Free Approach in Resonant Tunneling Systems}, \journal{Phys. Rev. B}{69}{195318}{2004}.
\bibitem{MyöhEtAl} P. Myoh{\"a}nen, A. Stan,  G. Stefanucci, and R. van Leeuwen, {\em A many-body approach to quantum transport dynamics: Initial
correlations and memory effects}, \journal{Europhys. Lett.}{84}{67001}{2008}.}
\bibitem{Albert2018} V.V. Albert, {\em Lindbladians with multiple steady states: theory and applications}, arXiv:1802.00010[quant-ph] (2018).
{{} \bibitem{PerfEtAl}  E. Perfetto, G. Stefanucci, and M. Cini, {\em Correlation-Induced Memory Effects in Transport Properties of Low-Dimensional Systems}, \journal{Phys. Rev. Lett.}{105}{156802}{2010}.
\bibitem{MatAnd} K. A. Matveev, and A. V. Andreev, {\em Equilibration of Luttinger Liquid and Conductance of Quantum Wires}, \journal{Phys. Rev. Lett.}{107}{056402}{2011}.
\bibitem{BuchEtAl} M. Buchhold, M. Heyl, and S. Diehl, {\em Prethermalization and Thermalization of a Quenched Interacting Luttinger Liquid}, \journal{Phys. Rev. A}{94}{013601}{2016}.}
\bibitem{Neqinventors} J. Schwinger, \journal{J. Math. Phys. (N.Y.)}{2}{407}{1961}; L.P. Kadanoff, G. Baym, {\em Quantum Statistical
Mechanics}, Benjamin, New York, 1962 R. P. Feynman, F. L. Vernon, \journal{Ann. Phys. (N.Y.)}{24}{118}{1963};
L.V. Keldysh, \journal{Sov. Phys. JETP}{20}{1018}{1965}.
{{} \bibitem{KarlEtAl}  D. Karlsson, R. van Leeuwen, Y. Pavlyukh,  E. Perfetto, and G. Stefanucci, {\em Fast Green’s function method for ultrafast electron-boson dynamics}, \journal{Phys. Rev. Lett.}{127}{036402}{2021}.}
\bibitem{Kubo1966} R. Kubo, {\em The fluctuation-dissipation theorem}, \journal{Rep. Prog. Phys.}{29}{255}{1966}.
\bibitem{BanEtAl2017} M. Ban, S. Kitajima, T. Arimitsu, and F.Shibata, {\em Linear response theory for open systems: Quantum master equation approach}, \journal{Phys. Rev. A}{95}{022126}{2017}.
\bibitem{PanEtAl2020} L. Pan, X. Chen,  Y. Chen, and H. Zhai, {\em Non-Hermitian linear response theory}, \journal{Nature Physics}{16}{767}{2020}.
\bibitem{AvronEtAl2012} J.E. Avron, M. Fraas, and G.M. Graf, {\em Adiabatic response for Lindblad dynamics}, \journal{J. Stat Phys.}{148}{800}{2012}.
\bibitem{WeiYan2011} J. H. Wei, and Y. Yan  {\em Linear response theory for quantum open systems},  arXiv:1108.5955 [cond-mat.mes-hall] (2011).
\bibitem{FeynVern} R. P. Feynman, and F. L. Vernon, {\em The theory of a general quantum system interacting with a linear dissipative system}, in \cite{Neqinventors}
\bibitem{Onsager} L. Onsager, {\em Reciprocal Relations in Irreversible Processes.}, \journal{Phys. Rev.}{37}{405}{1931}, \journal{Phys. Rev.}{38}{2265}{1931}.
\bibitem{GiesHajdu85} P. Gies and J. Hajdu, {\em Shubnikov-de haas effect in kinetic approximation}, \journal{Z. Phys. B}{60}{231}{1985}.
\bibitem{VIE90} O. Viehweger,  {\em A self-consistent approach to the localization-delocalization problem of the quantum Hall effect}, \journal{Z. Phys. B}{83}{45}{1991}.
\bibitem{Haag1967} R. Haag, N.M. Hugenholtz, and N. Winnink, {\em On the Equilibrium States in Quantum Statistical Mechanics}, \journal{Commun. math. Phys.}{5}{215}{1967}.
{{} \bibitem{Weiss} U. Weiss, {\em Quantum Dissipative Systems, 5th edn.}, World-Scientific, Singapore, (2021).}
\bibitem{Horn2009} K. Hornberger, {\em Introduction to decoherence theory}, \journal{Lect. Notes Phys.}{768}{221}{2009}.
{{} \bibitem{HoffEtAl}  N. M. Hoffmann, C. Sch{\"a}fer, N. Säkkinen, A. Rubio, H. Appel, and A.
Kelly, {\em Benchmarking Semiclassical and Perturbative Methods for Real-time
Simulations of Cavity-Bound Emission and Interference}, \journal{J. Chem. Phys.}{151}{244113}{2019}.}
\end{thebibliography}
\end{document}